\documentclass[twocolumn,iop,revtex4]{openjournal}

\pdfoutput=1
\usepackage{graphicx}
\usepackage{dcolumn}
\usepackage{bm}
\usepackage{amssymb,amsmath,bm}
\usepackage{color}
\usepackage[dvipsnames]{xcolor}
\usepackage[colorlinks,linkcolor=red,citecolor=blue,urlcolor=blue ]{hyperref}
\usepackage{multirow}
\usepackage[utf8]{inputenc}
\usepackage{balance}
\usepackage{enumitem}
\newcommand{\nv}{\hat{\bf n}}
\newcommand{\kalo}{Karhunen-Lo\`{e}ve~}
\newcommand{\cfh}{CFHTLenS~}

\begin{document}

\shorttitle{Sheer shear: weak lensing with one mode}
\title{Sheer shear: weak lensing with one mode}

\author{Emilio Bellini$^{1,*}$}
\thanks{$^*$emilio.bellini@physics.ox.ac.uk}
\author{David Alonso$^{2,1}$}
\author{Shahab Joudaki$^1$}
\author{Ludovic van Waerbeke$^3$}

\affiliation{$^{1}$Oxford Astrophysics, Department of Physics, Keble Road, Oxford, OX1 3RH, UK}
\affiliation{$^{2}$School of Physics and Astronomy, Cardiff University, The Parade, Cardiff, CF24 3AA, UK}
\affiliation{$^{3}$Department of Physics and Astronomy, University of British Columbia, Vancouver, BC V6T 1Z1, Canada}

\begin{abstract}
\noindent
3D data compression techniques can be used to determine the natural basis of radial eigenmodes that encode the maximum amount of information in a tomographic large-scale structure survey. We explore the potential of the \kalo decomposition in reducing the dimensionality of the data vector for cosmic shear measurements, and apply it to the final data from the \cfh survey. We find that practically all of the cosmological information can be encoded in one single radial eigenmode, from which we are able to reproduce compatible constraints with those found in the fiducial tomographic analysis (done with 7 redshift bins) with a factor of $\sim30$ fewer datapoints. This simplifies the problem of computing the two-point function covariance matrix from mock catalogues by the same factor, or by a factor of $\sim800$ for an analytical covariance. The resulting set of radial eigenfunctions is close to $\ell$-independent, and therefore they can be used as redshift-dependent galaxy weights. This simplifies the application of the \kalo decomposition to real-space and Fourier-space data, and allows one to explore the effective radial window function of the principal eigenmodes as well as the associated shear maps in order to identify potential systematics. We also apply the method to extended parameter spaces and verify that additional information may be gained by including a second mode to break parameter degeneracies. The data and analysis code are publicly available at \url{https://github.com/emiliobellini/kl_sample}.
\end{abstract}

\keywords{cosmology: large-scale structure of the Universe -- methods: data analysis}

\maketitle

\section{Introduction}
  Imaging galaxy surveys are one of the most promising paths towards understanding some of the main open questions in cosmology \citep{1999ApJ...522L..21H,2007MNRAS.381.1018A}, such as the precise nature of the energy components dominating the late-time expansion \citep{2003PhRvL..90i1301L}, the level of non-Gaussianity in the cosmic inhomogeneities induced by early-time physics \citep{2008PhRvD..77l3514D,2015PhRvD..92f3525A}, or the mass of neutrinos \citep{1998PhRvL..80.5255H}. To do so, these experiments combine the constraining power of several cosmological probes, such as Type-Ia supernovae, strong lenses, the distribution of galaxy clusters, the clustering of galaxies, and the weak gravitational lensing effect through the correlated distortion of galaxy shapes.
  
  Current and near-future photometric surveys, such as the Dark Energy Survey \citep{2017arXiv170801530D}, the Kilo-Degree Survey \citep{2018arXiv181206076H}, the Hyper Suprime-Cam \citep{2018arXiv180909148H}, Euclid \citep{2011arXiv1110.3193L} or the Large Synoptic Survey Telescope (LSST) \citep{2009arXiv0912.0201L} face a number of observational, statistical and computational challenges in order to extract the maximum amount of cosmological information from them without incurring significant systematic biases. Arguably the most critical challenge for these experiments is the correct treatment of the photometric redshifts \citep{2015APh....63...81N,2017MNRAS.465.1454H}. An object by object analysis would be expensive and the broad nature of the lensing kernel dilute the information contained by each galaxy. In addition, we lack perfectly accurate posterior redshift probability distributions for each object. This is why a so-called tomographic approach \citep{1999ApJ...522L..21H} has become the standard analysis path. In this case, the full galaxy sample is divided into photometric redshift (photo-$z$) slices, and cosmological constraints are extracted from the different cross-correlations between the 2D galaxy positions and shapes in all pairs of redshift bins. In this approach, the problem of obtaining and interpreting accurate posterior redshift distributions is replaced by the potentially simpler problem of estimating the redshift distribution of all objects in each redshift bin. The number and distribution of the different redshift bins used in the analysis is then determined as a compromise between the needs to maximize the amount of information recovered and to minimize the size of the final data vector (in the form of a collection of all auto- cross-bin two-point functions).
  
  In \cite{2018MNRAS.473.4306A}, we proposed a method to determine the optimal slicing for a photometric survey that recovers the maximum amount of information in the minimum number of radial modes. The method is based on the \kalo transform, and follows a formalism similar to other 3D data compression methods\footnote{In this paper, the use of the term ``3D data compression'' refers to the decomposition of a three-dimensional dataset, such as galaxy shear, into the set of modes spanned by a given set of radial basis functions.} proposed in the literature (e.g. \citealt{2003MNRAS.343.1327H,2007MNRAS.376..771K,2013arXiv1307.1307M,2014arXiv1403.5553K,2015PhRvD..92l3010L,2015A&A...578A..50A}). The resulting modes are in general scale- and redshift-dependent, and in the case of spectroscopic galaxy clustering can be related to the standard harmonic-Bessel transform that leads to the standard 3D power spectrum analysis, see a.g.~\cite{2018MNRAS.473.4306A}. In the case of cosmic shear, due to the radially cumulative nature of gravitational lensing, the \kalo transform can produce a significant reduction in the dimensionality of the final data vector, which has a direct impact on the computational complexity of estimating its covariance matrix (beyond guaranteeing an optimal recovery of cosmological information).

  Two main approaches have been used in the literature to attack the problem of covariance matrices for large-scale structure and weak lensing data. On the one hand, the sample covariance matrix can be computed by running a large number of simulations resembling the data, estimating the data vector in each of them, and estimating their sample variance. In order to recover a converged covariance that can be safely inverted, this requires running a number of simulations several times larger than the size of the data vector \citep{2013PhRvD..88f3537D,2016PhRvD..93f3524P} ($\mathcal{O}(10)$ times larger is often a good rule of thumb). Since the size of a complete set of tomographic cross-correlations scales as $\sim N_{\rm bin}^2$ with the number of tomographic bins, the computational complexity of estimating the sample covariance can grow very fast if a large number of bins is required in order to recover the cosmological information. On the other hand, it is also possible to compute analytical estimates of the covariance matrix. These analytical covariances \citep{2017MNRAS.470.2100K} are often faster to compute than the simulated ones, but they can also be computationally challenging, since it is often necessary to accurately account for the mode-coupling caused by the survey mask \citep[e.g. see][]{2019arXiv190611765G}, and real-space covariances can require expensive double Hankel transforms. Since all elements of the covariance matrix must be estimated individually, the complexity for analytical covariances scales with the square of the size of the data vector, or $\sim N_{\rm bin}^4$. The problem of estimating covariances can therefore benefit from efficient compression methods \citep{2017MNRAS.472.4244H}.
  
  As a proof of concept for the potential of this method, in this paper we apply it to the analysis of cosmic shear data from the Canada-France-Hawaii Telescope Lensing Survey (hereafter refered to as CFHTLenS, \citealt{2012MNRAS.427..146H}). Even if more up-to-date cosmic shear data are available, the choice of \cfh has been made mainly because other 3D data compression methods have been applied in the past \citep{2014MNRAS.442.1326K}, which makes possible a comparison between different methods. We build upon the tomographic analysis of \cite{2017MNRAS.465.2033J} and present the performance of our method in terms of optimal radial eigenmodes, compression factor, final constraints and goodness of fit. The paper is structured as follows: in Section \ref{sec:methods} we provide the theoretical background for weak lensing tomographic analyses and the \kalo transform as a data compression method. Section \ref{sec:data} then introduces the data used in our analysis, as well as the different data products we derive from it. Our main results are presented in Section \ref{sec:results}, which we discuss and summarize in Section \ref{sec:discussion}.
  
\section{Methods}\label{sec:methods}
  \subsection{Cosmic shear 2-point functions}\label{ssec:methods.wl2p}
    The cosmic shear field for galaxies observed in a given tomographic redshift bin $\alpha$ characterized by a redshift distribution $n_\alpha(z)$ is given by
    \begin{equation}
      (\gamma_1+i\gamma_2)_\alpha=-\eth\eth\left[\int_0^{\chi_H}d\chi\,\Phi(\chi\nv)\,q_\alpha(\chi)\right],
    \end{equation}
    where $\Phi(\chi\nv)$ is the gravitational potential in the lightcone towards a sky position with spherical coordinates $(\theta,\varphi)$ defined by the unit vector $\nv$, $\chi_H$ is the comoving radial distance to the horizon. The lensing kernel $q_\alpha$ in flat space becomes
    \begin{equation}\label{eq:lensing_kernel}
      q_\alpha(\chi)=\chi\int_{z(\chi)}^\infty dz\,\left(1-\frac{\chi}{\chi(z)}\right)n_\alpha(z).
    \end{equation}
    Here, $z$ is redshift, and $\eth$ is the spin-raising differential operator, which acts on a spin-$s$ quantity defined on the sphere $\,_sf(\nv)$ as\footnote{Note that the action of the first $\eth$ operator on the spin-0 quantity $\Phi$ makes it a spin-1 field, which affects the action on it of the second $\eth$ operator.}
    \begin{equation}
      \eth\,_sf\equiv-(\sin\theta)^s\left(\frac{\partial}{\partial\theta}+\frac{i}{\sin\theta}\frac{\partial}{\partial\varphi}\right)(\sin\theta)^{-1}\,_sf.
    \end{equation}
  
    The harmonic decomposition of a spin-2 quantity like $\gamma$ gives rise, in general, to both $E$- and $B$-modes \citep[e.g. see][]{1997PhRvD..55.1830Z}. Since gravitational lensing only gives rise to negligible $B$-modes, we will only concern ourselves with the $E$-mode component here (although our data analysis pipeline computes the $B$-mode component as a test of systematics).
    
    Since the gravitational lensing that shears the observed shape of a galaxy at redshift $z$ is caused by the collective effect of all the matter inhomogeneities along the line of sight to that redshift (manifested by the cumulative nature of the lensing kernels $q_\alpha$ in Eq.~\ref{eq:lensing_kernel}), the weak lensing signal in different redshift bins is usually strongly correlated. The effective number of degrees of freedom in a tomographic analysis therefore increases very slowly with the number of redshift bins, which motivates the use of radial data compression, as described in Section \ref{ssec:methods.kl}, to determine the optimal radial modes that saturate the information content of a given weak lensing survey.
    
    The two-point correlator of the weak lensing $E$-mode signal in two redshift bins $\alpha$ and $\beta$ at multipole $\ell$ can be related to the power spectrum of matter fluctuations $P(k,z)$ as \citep{2017MNRAS.472.2126K}
    \begin{equation}\label{eq:cell}
      S^{\alpha\beta}_\ell=\int \frac{d\chi}{\chi^2} \,\tilde{q}_\alpha(\chi)\,\tilde{q}_\beta(\chi)\,P\left(k=\frac{\ell+1/2}{\chi},z(\chi)\right)
    \end{equation}
    where $S$ stands for Signal, see Eq.~\ref{eq:totcell} below, and
    \begin{equation}\nonumber
      \tilde{q}_\alpha(\chi)\equiv F_\ell\,\frac{3H_0^2\Omega_M}{2\,a(\chi)}q_\alpha(\chi),\hspace{12pt}
      F_\ell\equiv\sqrt{\frac{(\ell+2)!}{(\ell-2)!}}\frac{1}{(\ell+1/2)^2},
    \end{equation}
    $a(\chi)$ is the scale factor at the comoving distance $\chi$, $H_0$ is the Hubble constant and $\Omega_M$ is the matter abundance at $z=0$. In deriving Eq.~\ref{eq:cell} we have used the Limber approximation\footnote{Note that the Limber approximation is accurate enough for weak lensing studies on the scales used here (\citealt{2017MNRAS.472.2126K}).} \citep{1953ApJ...117..134L,2004PhRvD..69h3524A}, and assumed a flat cosmology. Under the assumption that the observed shear is a combination of the true underlying signal and stochastic noise, the total power spectrum is also a sum of the signal and noise contributions:
    \begin{equation}\label{eq:totcell}
      C^{\alpha\beta}_\ell=S^{\alpha\beta}_\ell+N^{\alpha\beta}_\ell.
    \end{equation}
    We assume the noise power spectrum to be uncorrelated across bins (i.e. $N^{\alpha\beta}_\ell\propto\delta_{\alpha\beta}$), and we estimate it as described in Section \ref{sec:data}.
    
    The shear angular power spectrum can then be related to the real-space correlation functions $\xi_\pm$ as
    \begin{equation}
      \xi^{\alpha\beta}_\pm(\theta)=\int \frac{d\ell\,\ell}{2\pi}\,J_{2\mp2}(\ell\theta)\,C^{\alpha\beta}_\ell,
    \end{equation}
    where $J_\nu(x)$ is the cylindrical Bessel function of order $\nu$, and we have used the flat-sky approximation, which is appropriate for the small angular footprint covered by the \cfh survey, i.e.~$\sim 154\, {\rm deg}^2$.
    
    To estimate the theoretical predictions for all two-point functions used in this work we used the Core Cosmology Library (\textsc{CCL}) \citep{2018arXiv181205995C}, with the matter power spectrum computed by \textsc{CLASS} \citep{2011arXiv1104.2932L} using the version of \textsc{Halofit} by \cite{2012ApJ...761..152T}.

  \subsection{The \kalo transform and radial data compression}\label{ssec:methods.kl}
    Suppose we have a data vector ${\bf x}$ with dimension $N_s$, that we assume Gaussianly distributed with zero mean, and whose covariance ${\sf C}\equiv\langle {\bf x} {\bf x}^\dagger\rangle$ depends on a particular parameter $\theta$ we want to measure. It is possible to find a linear transformation ${\sf E}$ such that the elements of the transformed data vector ${\bf y}\equiv{\sf E}^\dagger\,{\bf x}$ have unit variance and are uncorrelated (i.e.~$\langle y_p y_q^*\rangle=\delta_{pq}$), and such that the first $m<N_s$ elements contain most of the information about $\theta$ \citep{1997ApJ...480...22T,1995PhRvL..74.4369B,1996ApJ...465...34V}. The columns of ${\sf E}$, which we label ${\bf e}_p$, are the eigenvectors of the following generalized eigenvalue problem:
    \begin{equation}\label{eq:kl_general}
      \partial_\theta{\sf C}\,{\bf e}_p=\lambda_p\,{\sf C}\,{\bf e}_p,
    \end{equation}
    where $\partial_\theta$ denotes the partial derivative with respect to $\theta$. The resulting linear transformation is the so-called \kalo (KL) transform, which can be used to simplify the analysis for one particular parameter.
    
    In a more general situation, we would like to measure several parameters simultaneously. Although under certain circumstances it may be possible to find a set of KL eigenmodes that maximize the overall information content, as described in \cite{1997ApJ...480...22T}, a simpler and popular approach is to instead maximize the information on the overall amplitude of the signal. For concreteness, let us decompose our data into signal and noise contributions ${\bf x}={\bf s}+{\bf n}$, and let us introduce a fictitious amplitude $\alpha$ with fiducial value 1 such that in general ${\bf x}=\alpha{\bf s}+{\bf n}$. Under the assumption that signal and noise are uncorrelated, and redefining $2/(2-\lambda_p)\rightarrow\lambda_p$, we can rewrite the generalized eigenvalue problem in Eq.~\ref{eq:kl_general} as
    \begin{equation}
      {\sf C}{\bf e}_p=\lambda_p{\sf N}\,{\bf e}_p,
    \end{equation}
    where ${\sf C}={\sf S}+{\sf N}$ is the covariance matrix of ${\bf x}$ as a sum of the signal and noise covariances. Using the Cholesky decomposition of the noise covariance ${\sf N}={\sf L}{\sf L}^\dagger$, and defining $\tilde{\bf e}_p\equiv{\sf L}^\dagger{\bf e}_p$, the previous equation can be rewritten as a standard eigenvalue problem for $\tilde{\bf e}_p$:
    \begin{equation}\label{eq:kl_eigenvalue}
      \left[{\sf L}^{-1}{\sf C}\,({\sf L}^{-1})^\dag\right]\,\tilde{\bf e}_p=\lambda_p\tilde{\bf e}_p.
    \end{equation}
    Solving for $\tilde{\bf e}_p$, the new data vector can then be computed as $y_p=\tilde{\bf e}_p^\dag\,{\sf L}^{-1}{\bf x}$. The covariance of the new data vector is given by the eigenvalues $\langle y_py^*_{p'}\rangle=\lambda_p\delta_{pp'}$, and the transformed noise covariance is simply the identity.

    As described in \cite{2018MNRAS.473.4306A}, this formalism can be used to determine the optimal set of radial modes in a given photometric redshift survey. In this case, our data vector ${\bf x}$ is the vector of $E$-mode harmonic coefficients of the cosmic shear field in different redshift bins for a fixed $(\ell,m)$, and ${\sf S}$, ${\sf N}$ and ${\sf C}$ are the signal, noise and total power spectra described in the previous section at the corresponding value of $\ell$. The KL eigenmodes $e^\alpha_{p,\ell}$ are therefore scale-dependent functions in general. However, as shown in \cite{2018MNRAS.473.4306A}, this scale dependence is very mild for cosmic shear data, and in this case the eigenmodes can be thought of as redshift-dependent galaxy weights that can be used to generate the set of most informative shear maps. We will explore this in detail in Section \ref{sec:results}. In this case, each of the KL modes can be interpreted as a shear map associated with an effective redshift distribution given by $\tilde{n}_p(z)=e^\alpha_p\,n_\alpha(z)$.

\section{Data}\label{sec:data}
  \subsection{The \cfh lensing sample}\label{ssec:data.sample}
  We use data from the \cfh survey\footnote{\url{http://www.cfhtlens.org/}} \citep{2012MNRAS.427..146H}. Their analysis performed weak lensing data processing with \textsc{THELI} \citep{2013MNRAS.433.2545E}, shear measurement with \textsc{lensfit} \citep{2013MNRAS.429.2858M}, and photometric redshift measurement with PSF-matched photometry \citep{2012MNRAS.421.2355H}. A full systematic error analysis of the shear measurements in combination with the photometric redshifts is presented in \cite{2012MNRAS.427..146H}, with additional error analyses of the photometric redshift measurements presented in \cite{2013MNRAS.431.1547B}.
  
  \cfh is a deep multi-colour survey that spans $154\,{\rm deg}^2$ of the sky divided in four fields -- labelled W1, W2, W3 and W4. The full \cfh catalogue contains $\sim 2\times 10^7$ objects (precisely 22,652,728) that have been classified as ``galaxy'', ``star'' or ``bad fit object'' by the \texttt{lensfit} algorithm (see \citealt{2013MNRAS.429.2858M} for details). Furthermore, some of the objects are in masked regions of the sky due to observational systematics, or their position is in MegaCam fields that did not pass the lensing residual systematics effects, as described in \cite{2012MNRAS.427..146H}. Thus, in any shear analysis that uses \cfh data, the catalogue has to be cleaned from objects that are classified as stars or lie on unreliable sky regions. In detail, the filters we used in order to obtain a clean catalogue are:
  \begin{itemize}
    \item \texttt{MASK}$=0$: to select only unmasked objects;
    \item \texttt{weight}$>0$: objects that have been measured with strictly positive weight;
    \item \texttt{star\_flag}$=0$: remove stars and bad fits, and keep only galaxies;
    \item remove objects with positions in the $1\, {\rm deg}^2$ camera MegaCam fields that do not pass the lensing residual systematics effects. For a list of those fields see \cite{2013MNRAS.433.2545E}.
  \end{itemize}
  After using these filters, the clean catalogue reduces to 4,180,003 objects, which we use to estimate the lensing two-point statistics.
  
\begin{figure}
      \centering
      \includegraphics[width=0.9\columnwidth]{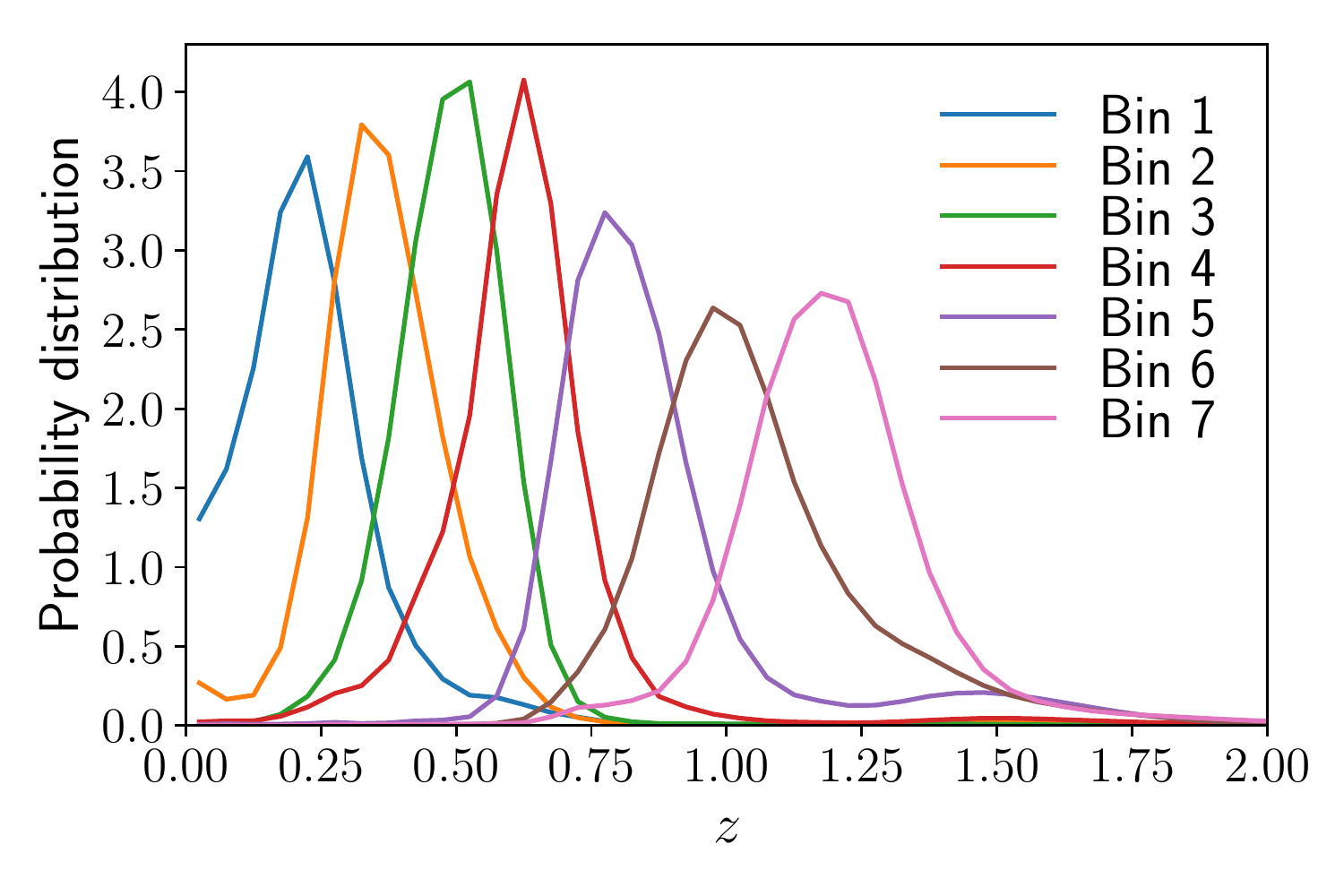}\vspace{-5pt}
      \caption{Stacked probability distribution of each unmasked galaxy per redshift bin \citep{2017MNRAS.465.2033J}. Each colour refers to a different bin, defined in Table \ref{tab:z_bins}.}\label{fig:photoz}
    \end{figure}

  The ellipticities of each source, as well as their additive ($c_i$) and multiplicative ($m_i$) correction biases and an approximately optimal inverse variance weight ($w_i$) are obtained using the Bayesian model fitting software \textsc{lensfit} (\citealt{2013MNRAS.429.2858M}). Photometric redshifts are derived from the Bayesian photometric redshift code \textsc{BPZ} (\citealt{2000ApJ...536..571B}), which returns a full redshift probability distribution $p(z)$ for each object. As in \cite{2017MNRAS.465.2033J}, we split the sample into 7 tomographic bins, using the peak of the redshift distribution of each galaxy ($z_B$) as a redshift bin label. Table \ref{tab:z_bins} shows the redshift bins chosen along with the effective density of galaxies in each bin, defined as
    \begin{equation}\label{eq:n_eff}
      n_{\rm eff} = \frac{\left(\sum_i w_i\right)^2}{A_{\rm eff}\left(\sum_i w_i^2\right)}\,.
    \end{equation}
    Here the effective area ($A_{\rm eff}$) per field is [41.0, 11.6, 24.7, 12.2] deg$^{-2}$ for W1-4 respectively. The redshift distribution for each tomographic bin was obtained, as in \cite{2017MNRAS.465.2033J}, by stacking the photo-$z$ $p(z)$ of all unmasked objects in that bin. The resulting normalized redshift distributions are shown in Figure \ref{fig:photoz}.

    \begin{table}
      \begin{center}
        \begin{tabular}{cc|ccccc}
          \hline
          Bin No. & $z_B$ & \multicolumn{5}{c}{$n_{\rm eff}$ [arcmin$^{-2}$]}\\
          &  & W1 & W2 & W3 & W4 & TOT\\
          \hline\hline
          1 & $0.15$ -- $0.29$ & $1.41$ & $1.38$ & $1.39$ & $1.37$ & $1.40$\\
          2 & $0.29$ -- $0.43$ & $1.18$ & $1.23$ & $1.18$ & $1.12$ & $1.18$\\
          3 & $0.43$ -- $0.57$ & $1.16$ & $1.17$ & $1.30$ & $1.19$ & $1.21$\\
          4 & $0.57$ -- $0.70$ & $1.70$ & $1.57$ & $1.84$ & $1.80$ & $1.73$\\
          5 & $0.70$ -- $0.90$ & $2.69$ & $2.58$ & $2.91$ & $2.87$ & $2.76$\\
          6 & $0.90$ -- $1.10$ & $1.96$ & $1.49$ & $1.91$ & $1.67$ & $1.85$\\
          7 & $1.10$ -- $1.30$ & $1.07$ & $0.89$ & $0.94$ & $0.83$ & $0.98$\\
          \hline
        \end{tabular}
        \caption{Redshift bins used in this analysis along with the effective density $n_{\rm eff}$ per bin and field. The last column shows the total $n_{\rm eff}$ per bin.}\label{tab:z_bins}
      \end{center}
    \end{table}  
  
  \subsection{Masks and maps}\label{ssec:data.masks}
    In this section we describe in detail the method used to generate masks and maps of shear measurements, which are then used to compute the power spectra. As mentioned, the relative small area covered by \cfh allows to use a flat-sky approximation. In this case, masks and maps are 2D arrays of pixels (one array for each field). In each pixel, the masks quantify the quality of the \cfh measurements in a given sky area, while the maps store the value of the two shear components estimated by averaging over the ellipticities of all sources in the pixel.

\begin{figure}
      \centering
      \includegraphics[width=0.9\columnwidth]{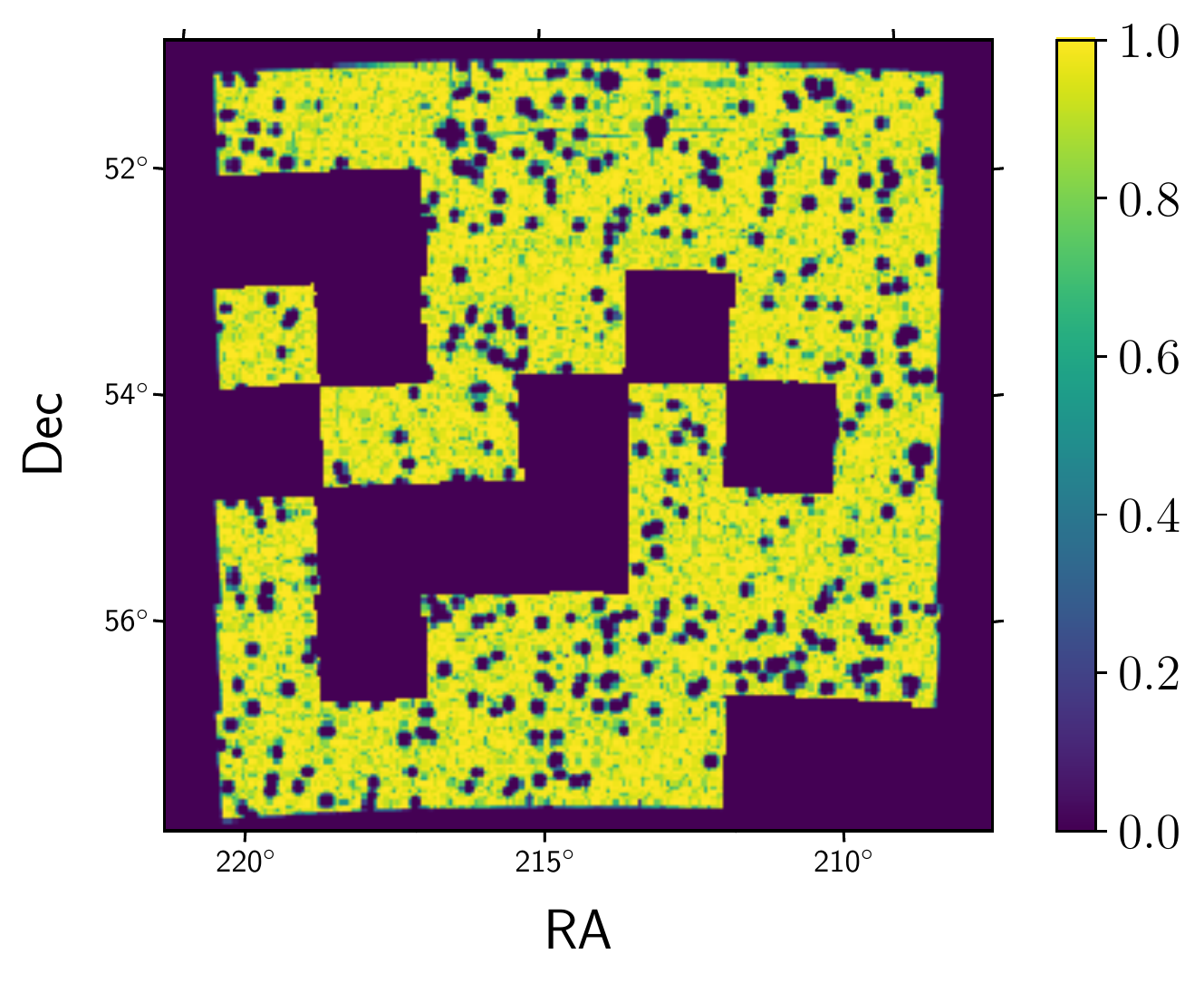}\\
      \includegraphics[width=0.9\columnwidth]{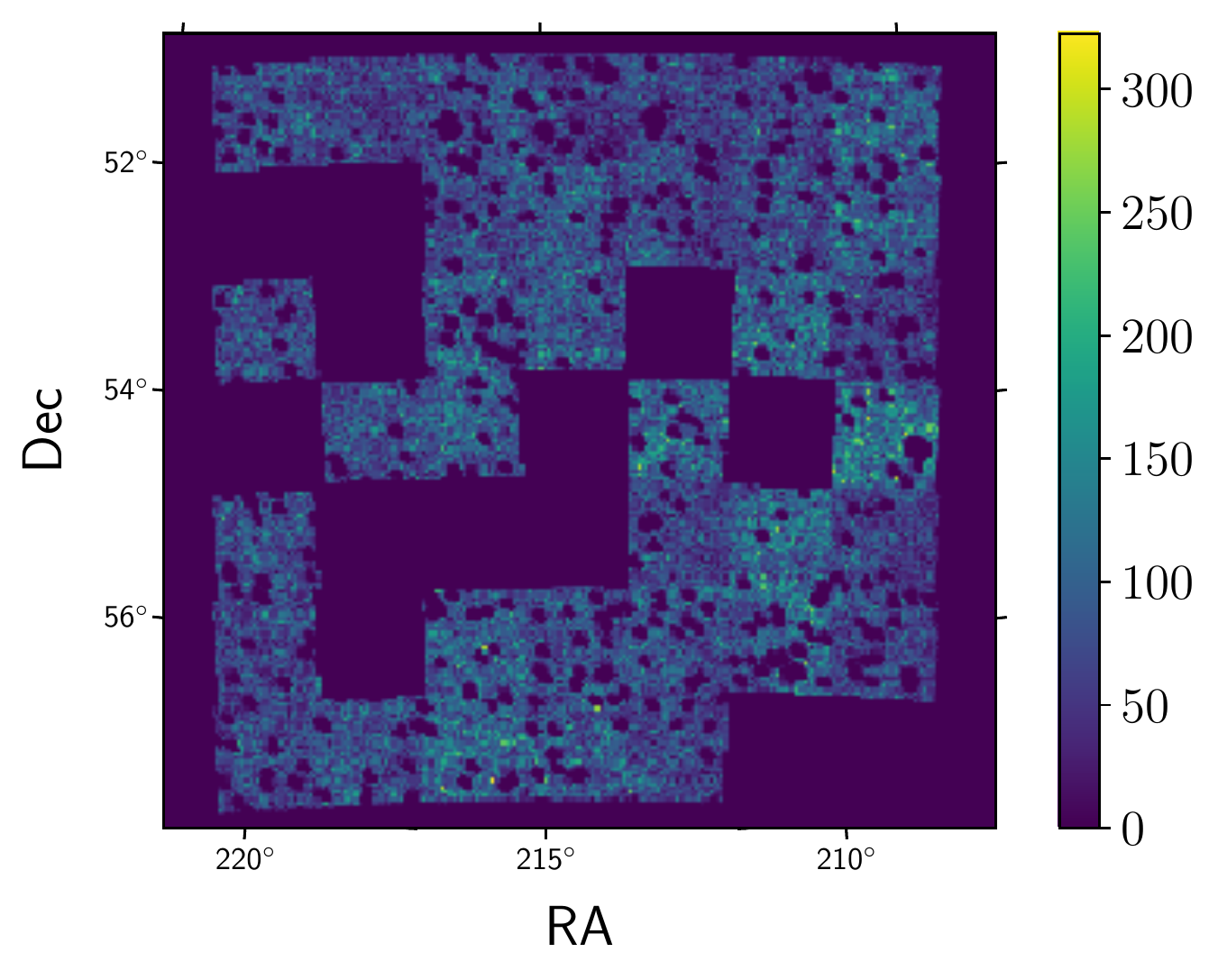}\vspace{-5pt}
      \caption{Example of masks used in this analysis for the W3 field. The top figure is the mask without including the \textsc{lensfit} weights. The bottom panel is the weights map for the sixth redshift bin.}\label{fig:mask}
    \end{figure}

    To define our sky mask, we start from a set of individual sky masks for each of the 4 \cfh fields. These have a resolution of 1 arcsecond and allow us to remove problematic areas, such as those outside the \cfh footprint, around stars and stellar halos, significant object overdensities, flagged pixels, and others. The full description of the masked areas can be found in Table B2 of \cite{2013MNRAS.433.2545E}. While in \cite{2013MNRAS.433.2545E} the authors mention that it is possible to consider as unmasked areas those that lie in large masks around stars and stellar haloes ({\tt mask} value of $1$), here we adopt a more conservative approach, and consider only areas with {\tt mask} value of $0$. We also remove by hand all pixels falling inside any of the pointings identified by \cite{2012MNRAS.427..146H} as being systematics dominated (corresponding to a total of $\sim25\%$ of the available area). Once all sources in masked pixels are removed from the data, we downgrade the resolution of this binary mask to a pixel size of 2 arcminutes, while preserving the information in the finer map in terms of the fractional area masked in each larger pixel. We chose this resolution to guarantee that the majority of the unmasked pixels will be populated by, on average, four galaxies in each redshift bin. Finally, to optimally weight each pixel when computing the angular power spectra, for each field and redshift bin we define the weights map used to compute the power spectrum (see Section \ref{ssec:data.2pt}) to be proportional to the sum of the \textsc{lensfit} weights of all sources in each pixel. Figure \ref{fig:mask} shows the mask and weights map used in this analysis for a particular field (W3). The top panel shows the 2 arcminutes mask obtained by downgrading the original 1 arcsec binary mask. This is the mask that has been used to calculate the area of each field used in Eq.~\ref{eq:n_eff}. The bottom panel represents the weights map described above, used by the power spectrum estimator. It is worth noting that the weights map is therefore different for each redshift bin, and the associated mode-coupling matrix (described in the next section) must be computed independently for each pair of bins.

\begin{figure}
      \centering
      \includegraphics[width=0.9\columnwidth]{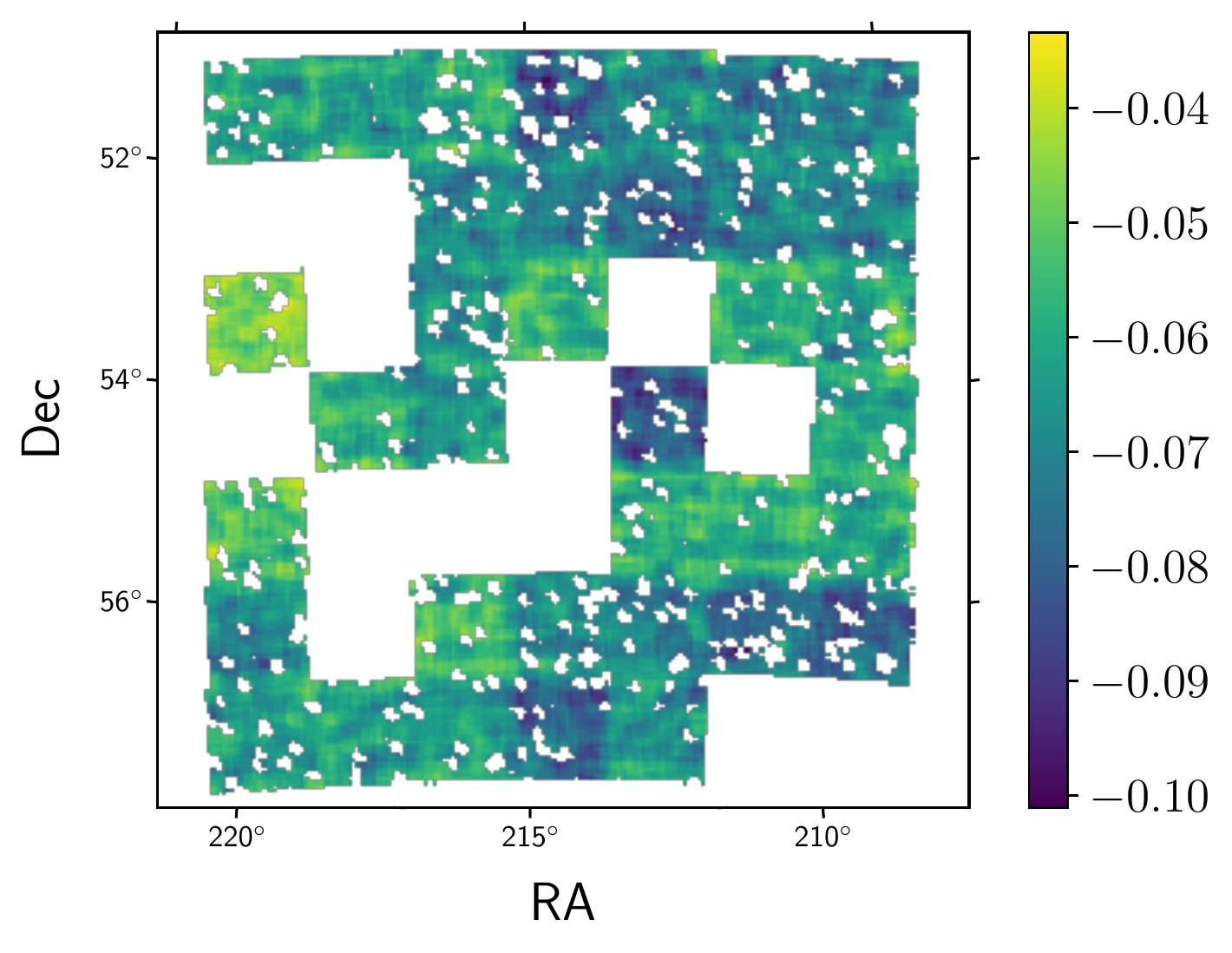}\vspace{-5pt}
      \caption{Map of the multiplicative corrections ($m_p$) for the W3 field and the sixth redshift bin.}\label{fig:mult_corr}
    \end{figure}

\begin{figure}
      \centering
      \includegraphics[width=0.9\columnwidth]{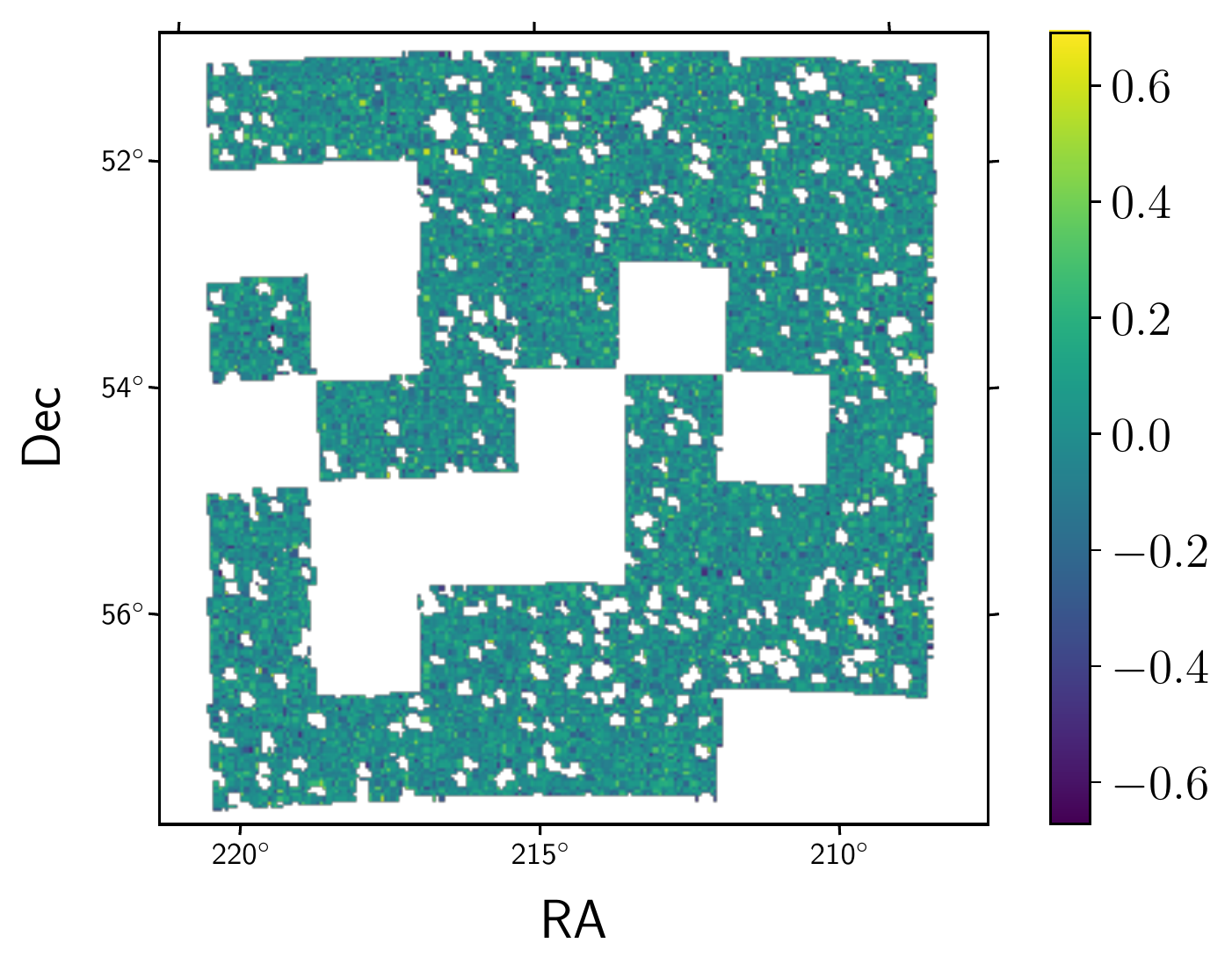}\\
      \includegraphics[width=0.9\columnwidth]{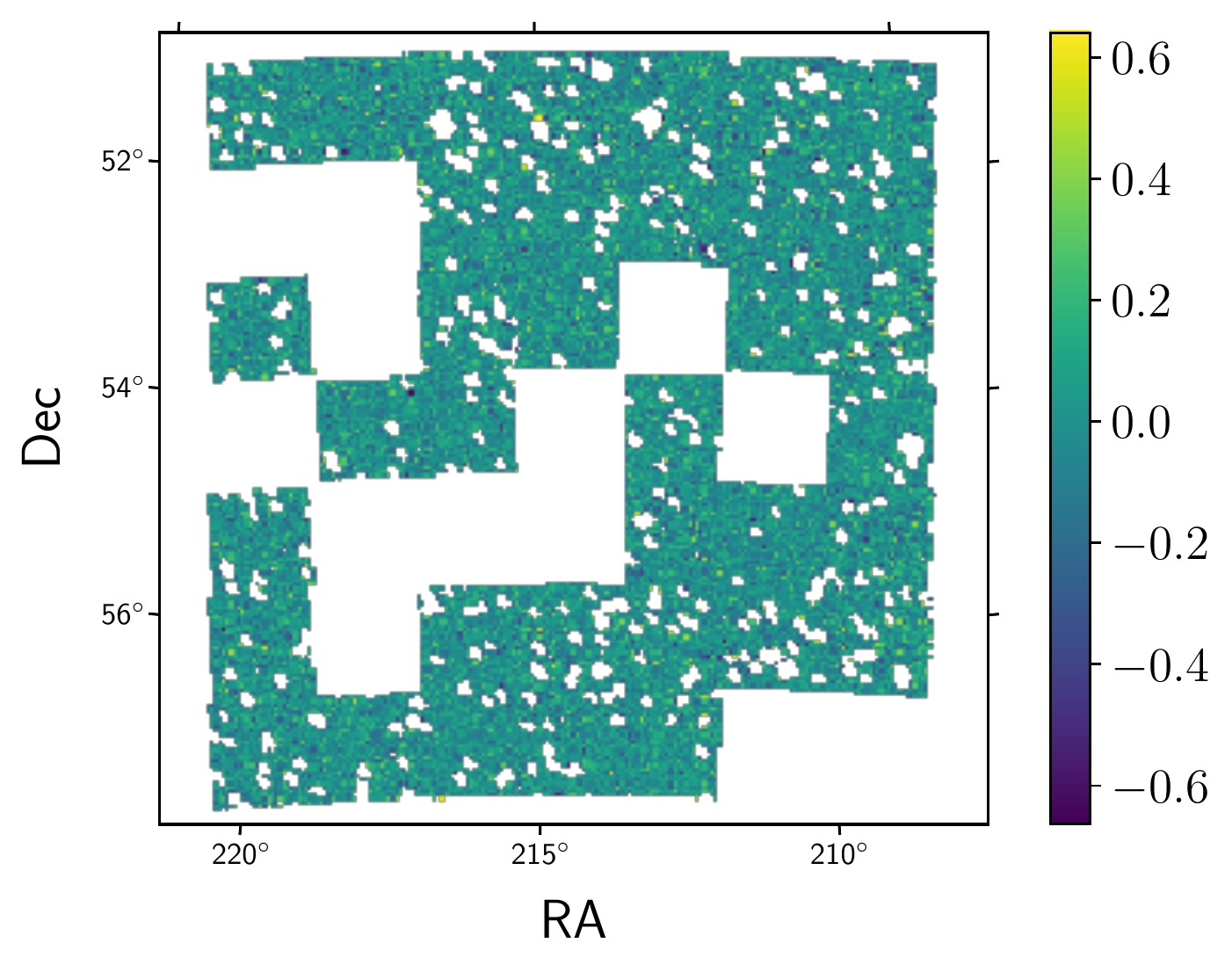}\vspace{-5pt}
      \caption{Maps used in this analysis for the W3 field and sixth redshift bin. The top panel is the map of $\gamma_1(\nv)$ while the bottom panel is the map of $\gamma_2(\nv)$.}\label{fig:map}
    \end{figure}

    To compute power spectra we also need to produce maps of the two shear components $\gamma_1(\nv)$ and $\gamma_2(\nv)$. A first estimate of these maps is made as a weighted average of the ellipticity components of all sources falling within each pixel corrected for their measured additive and multiplicative shape measurement bias $c_{i,g}$ and $m_p$
    \begin{equation}
      \tilde{\gamma}_i(\nv_p)=\frac{\sum_{g\in p}w_g \left(e_{i,g}-c_{i,g}\right)}{\left(1+m_p\right) \sum_{g\in p}w_g}\,,
    \end{equation}
    where $g$ stands for galaxy and $p$ for pixel. While the additive correction can be applied on a object by object basis (noting that $c_1=0$ in the \cfh analysis), $m$ must be estimated by averaging over a large number of sources to avoid a large bias in the final maps \citep{2013MNRAS.429.2858M}. Thus, we make a map of $m_p$ for each field and redshift bin by averaging, for a given pixel $p$, over the multiplicative bias parameter of all sources lying within a square of size 10 arcmin centred around $p$. In Figure \ref{fig:mult_corr} we show the multiplicative correction map for the W3 field and the sixth redshift bin (the bin with the highest shear signal-to-noise ratio). It is possible to notice that this gives up to a $\sim10\%$ correction to the shear measurements. Finally, Figure \ref{fig:map} shows the shear maps used in this analysis for the W3 field and the sixth redshift bin, with $\gamma_1$ and $\gamma_2$ shown in the top and bottom panels respectively.

  \subsection{Two-point functions}\label{ssec:data.2pt}
    \begin{figure*}
      \centering
      \includegraphics[width=0.9\textwidth]{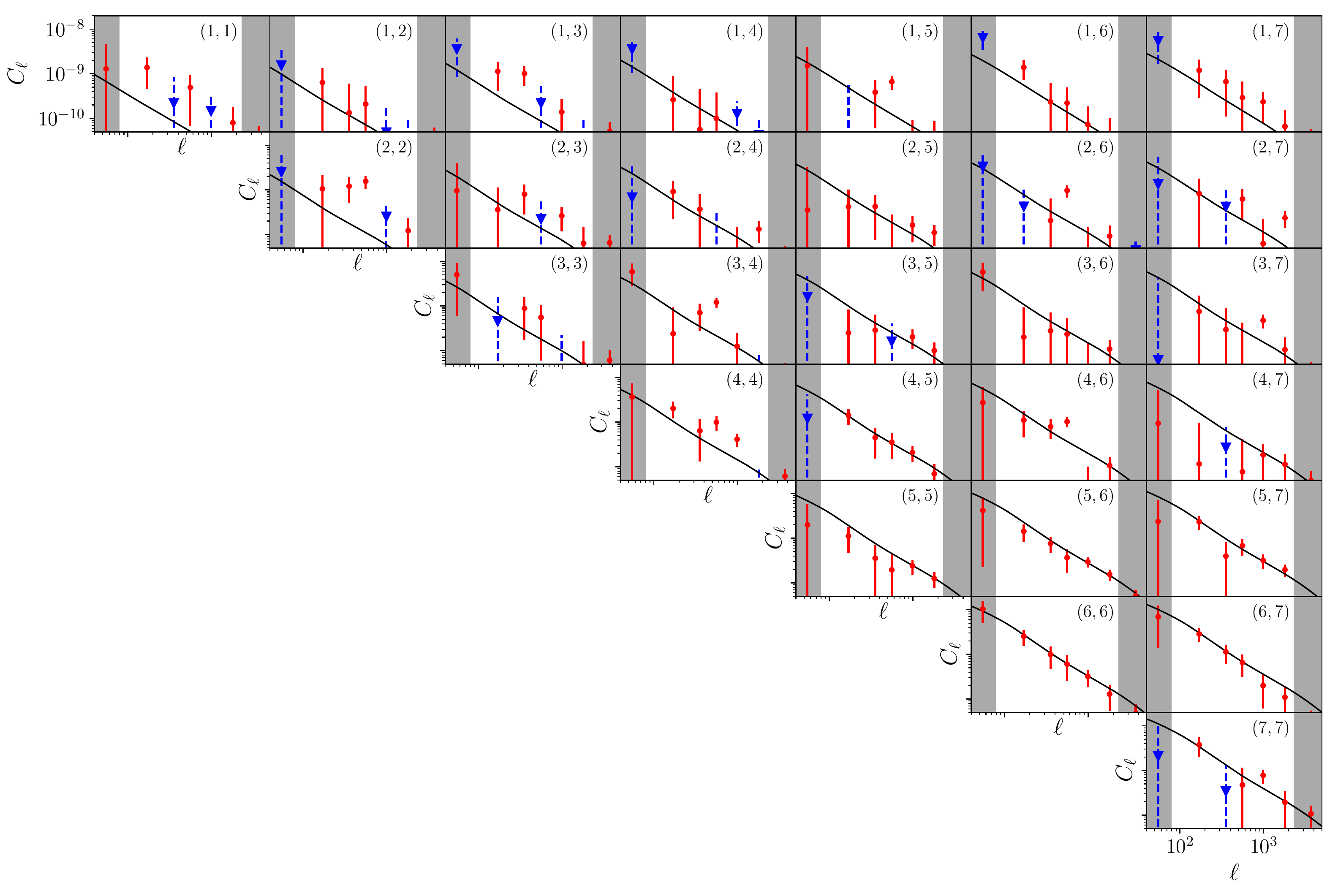}\vspace{-5pt}
      \caption{Measurements of the E-mode power spectra estimated from \cfh data. Red points and red solid error bars indicate positive $C_\ell$, while blue triangles and dashed error bars refer to negative $C_\ell$. Solid black lines show the theoretical predictions for the best fit cosmology of \citep{2017MNRAS.465.2033J}. Gray shaded areas represent $\ell$-bandpowers that have been discarded in this analysis.}\label{fig:cls_data}
    \end{figure*}
    We estimate all $EE$ and $BB$ cross-power spectra between the 7 different redshift bins using a pseudo-$C_\ell$ estimator as implemented in \textsc{NaMaster} \cite{2018arXiv180909603A}. We describe the estimator briefly here and refer the reader to the original paper for further details.
    
    The pseudo-$C_\ell$ estimator attempts to make an unbiased measurement of the power spectrum of a masked field by analytically computing the coupling between modes induced by that mask. Mathematically, let $f^v(\nv)=v(\nv)\,f(\nv)$ be the observation of an underlying true field $f$ through a multiplicative weights map $v$. Minimally, this weights map will be a binary mask removing regions where the field has not been observed ($v=0$) leaving only the observed pixels ($v=1$). More optimally, the weights map should be proportional to the local inverse noise variance, downweighting or upweighting different regions depending on their noise properties. In our analysis, the weights map is given by the sum of \textsc{lensfit} weights in each pixel.
    
    \begin{table}
    \vspace{5pt}
    \begin{center}
    \begin{tabular}{ccc}
    \hline
    No. & $\ell$-range & Used\\
    \hline
    1 & 30 -- 80 & \\
    2 & 80 -- 260 & \checkmark\\
    3 & 260 -- 450 & \checkmark\\
    4 & 450 -- 670 & \checkmark\\
    5 & 670 -- 1310 & \checkmark\\
    6 & 1310 -- 2300 & \checkmark\\
    7 & 2300 -- 5100 & \\
    \end{tabular}
    \caption{We show the $\ell$-bandpowers used in this analysis. The first column indicates the number of the bandpower, the second one the $\ell$-range that each bandpower covers and the last column indicates which bandpowers were actually used to get cosmological constraints.}\label{tab:bandpowers}
    \end{center}
    \end{table}
    Due to the convolution theorem, a direct estimate of the power spectrum from the masked fields $\tilde{C}^{fg}_\ell$ ignoring their masks, will in general yield a mode-coupled version of the underlying true power spectrum
    \begin{equation}
      \left\langle\tilde{C}^{fg}_\ell\right\rangle\equiv\left\langle\frac{1}{2\ell+1}\sum_{m=-\ell}^{\ell}{\rm Re}(f^v_{\ell m}g^{w*}_{\ell m})\right\rangle=\sum_{\ell'} M^{vw}_{\ell\ell'}\,C^{fg}_{\ell'},
    \end{equation}
    where the mode-coupling matrix $M^{vw}_{\ell\ell'}$ depends on the masks $v$ and $w$, and can be estimated analytically. The mode-coupling matrix is usually ill-conditioned and cannot be directly inverted to yield an unbiased estimate of $C^{fg}_\ell$. The usual approach is then to bin the pseudo-$C_\ell$ into bandpowers:
    \begin{equation}
      \tilde{\cal C}^{fg}_k=\sum_{\ell=\ell_{\rm min}^k}^{\ell_{\rm max}^k}\tilde{C}^{fg}_\ell.
    \end{equation}
    For this work we use the same $\ell$ bins chosen by \cite{2016MNRAS.456.1508K}\footnote{Note that, in order to obtain stronger cosmological constraints, in this paper we use one additional bandpower w.r.t.~\cite{2016MNRAS.456.1508K}, namely the 6th, i.e.~$\ell=1310$~--~$2300$.} and summarized in Table \ref{tab:bandpowers}. This choice was motivated by the need to capture the most relevant power spectrum features while staying within the mildly non-linear regime. For sufficiently wide bandpowers, the corresponding binned coupling matrix 
    \begin{equation}
      {\cal M}^{vw}_{kq}\equiv\sum_{\ell=\ell_{\rm min}^k}^{\ell_{\rm max}^k}\sum_{\ell'=\ell_{\rm min}^q}^{\ell_{\rm max}^q}M^{vw}_{\ell\ell'}
    \end{equation}
    becomes invertible, and the final estimator takes the form
    \begin{equation}\label{eq:pcl}
      {\cal C}^{fg}_k=\left({\cal M}^{vw}\right)^{-1}_{kq}\sum_{\ell=\ell_{\rm min}^q}^{\ell_{\rm max}^q}(\tilde{C}^{fg}_\ell-\tilde{N}^{fg}_\ell),
    \end{equation}
    where $\tilde{N}^{fg}_\ell$ is an estimate of the noise pseudo-power spectrum (the so-called ``noise bias''). This estimator, described here for scalar fields on the sphere, can be easily generalized to spin-2 quantities and flat skies, which we use in this work. We refer the reader to \cite{2018arXiv180909603A} for further details.
    \begin{figure*}
      \centering
      \includegraphics[width=0.9\textwidth]{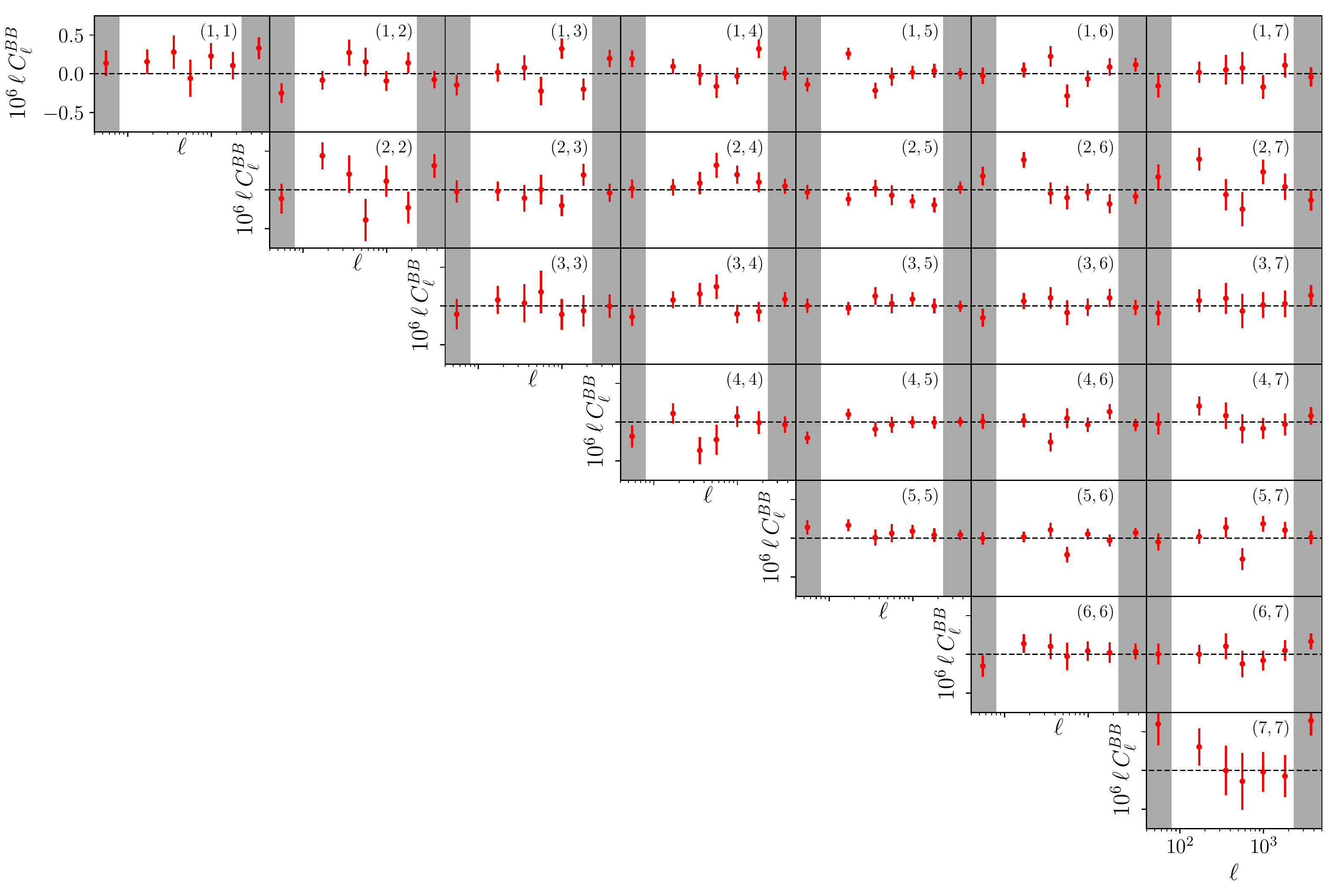}\vspace{-5pt}
      \caption{Same as Fig.~\ref{fig:cls_data} for the $B$-mode power spectra. Here, the grey areas represent the $\ell$-bandpowers that were excluded in the analysis, where a detection of $B$-modes should not be taken into account. We do not find statistically significant $B$-modes, with a PTE of $\sim10\%$ for the measurements shown. Gray shaded areas represent $\ell$-bandpowers that have been discarded in this analysis.}\label{fig:cls_data_BB}
    \end{figure*}
    
    Finally, as in \cite{2018arXiv180909148H}, we estimate the noise bias in Eq.~\ref{eq:pcl} by averaging over the power spectrum measured in 1000 noise realizations, generated by randomly rotating the ellipticities of all sources in the catalogue.
    
    This procedure allows us to estimate the power spectra in each of the 4 \cfh fields. The combined power spectra are then computed as an inverse-variance sum of the per-field power spectra:
    \begin{equation}
      {\cal C}_{\rm tot}=\mathsf{\Sigma}_{\rm tot}\,\sum_{f=1}^4\mathsf{\Sigma}^{-1}_f{\cal C}_f,
      \hspace{12pt}
      \mathsf{\Sigma}_{\rm tot}^{-1}=\sum_{f=1}^4 \mathsf{\Sigma}_f^{-1}.
    \end{equation}
    Here ${\cal C}_f$ is the vector of all power spectra computed in field $f$, $\mathsf{\Sigma}_f$ is the power spectrum covariance matrix in field $f$, described in the next section, and ${\cal C}_{\rm tot}$ and $\mathsf{\Sigma}_{\rm tot}$ are the combined power spectra and covariance matrix.

    The resulting measurements of the $E$-mode power spectra estimated from the \cfh data are shown in Figure~\ref{fig:cls_data}, where the error bars are computed from the covariance matrix described in Section~\ref{ssec:data.covar}. We find good agreement with the best-fit cosmological model of \cite{2017MNRAS.465.2033J} (solid lines). It is worth noting that this is the first time that a pseudo-$C_\ell$ analysis of these data has been carried out. \cite{2017MNRAS.465.2033J} used real-space correlations measured on these same redshift bins, while \cite{2016MNRAS.456.1508K} used an optimal quadratic estimator on two redshift bins that would be computationally impractical for a larger number of subsamples. After implementing the scale cuts shown in Figure~\ref{fig:cls_data}, the final data vector has 140 elements.
    
    We use three sanity checks to validate the power spectrum estimation. First, running our pipeline on the set of 2000 Gaussian simulations used to calculate the covariance matrix (described in Section \ref{ssec:data.covar}), we verify that we are able to recover the input power spectra with no appreciable bias (this was also thoroughly verified in \citealt{2018arXiv180909603A}). Secondly, we estimate the $B$-mode power spectrum of all pairs of redshift bins in the \cfh data. Since the cosmological signal is not expected to yield a detectable $B$-mode signal, the presence of $B$-modes can be used as a diagnostic for unknown systematics. Within the range of scales used in this analysis we do not detect significant $B$-modes, with a probability to exceed (PTE\footnote{The PTE here is the probability of the $\chi^2$ to be larger than the value found in the data.}) associated with the measurements shown in Figure~\ref{fig:cls_data_BB} of $\sim10\%$. Lastly, with our pipeline we replicated the analysis done in \cite{2016MNRAS.456.1508K}, and the resulting $E$- $B$-modes match perfectly. Other $B$-mode analyses of the  \cfh data have been carried out in \cite{2014MNRAS.442.1326K}, \cite{2017MNRAS.464.1676A} and \cite{2017MNRAS.466.3272A}.
    
    Although our main analysis uses Fourier-space observables, we also present results for the measurements of the correlation function made by \cite{2017MNRAS.465.2033J}, in order to make a direct comparison with those results. The measurements correspond to the 7 tomographic bins from $z=0.15-1.3$ and 7 angular bins from $\theta=1-120$~arcmin, which post-masking results in a data vector of 280 elements. The correlation functions $\xi_{\pm}$ were computed for each pair of redshift bins in each of the four fields using the \textsc{athena} software\footnote{\url{http://www.cosmostat.org/software/athena}} \citep{2014ascl.soft02026K}. These measurements include the same additive shear correction as in the Fourier space analysis, and moreover include a weighted ensemble average multiplicative shear correction following \cite{2012MNRAS.427..146H} and \cite{2013MNRAS.429.2858M}. The patch-dependent measurements were subsequently averaged by the number of galaxy pairs returned by \textsc{athena}.
  
  \subsection{Covariance matrix}\label{ssec:data.covar}
    \begin{figure}
      \centering
      \includegraphics[width=0.49\textwidth]{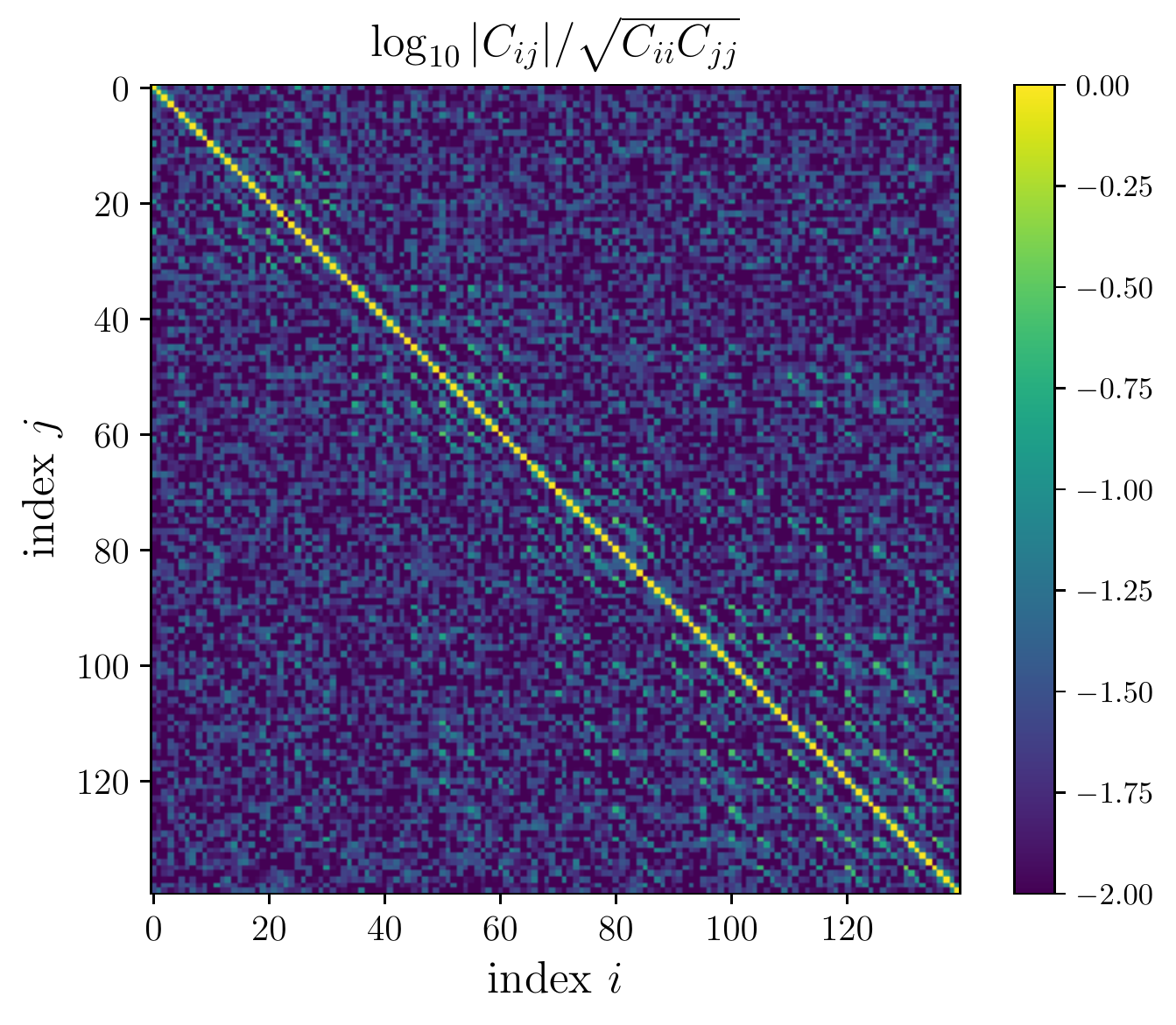}\vspace{-5pt}
      \caption{Correlation matrix for the $E$-mode power spectra estimated from a set of 2000 Gaussian simulations. The bin pairs are shown in row-major order (i.e. $((1,1),(1,2),...,(1,7),(2,2),(2,3),...,(6,7),(7,7))$).}\label{fig:covar_cls}
    \end{figure}
    To estimate the covariance matrix of the power spectrum measurements we make use of 2000 Gaussian simulations of the 4 fields. Each simulation is generated in two steps:
    \begin{enumerate}
      \item We generate a Gaussian realization of the expected cosmic shear signal $\gamma$. To do so, we draw Gaussian realizations of the Fourier-space $E$-modes with a variance given by the $E$-mode power spectrum predicted by the best-fit cosmological parameters of \cite{2017MNRAS.465.2033J}, and no $B$-modes. These are then transformed into real space to produce a shear map. Note that the Gaussian Fourier coefficients are not drawn independently for each redshift bin, but fully preserve the expected correlation between them.
      \item We then interpolate the Gaussian shear field to the position of each source in the original catalogue. The total ellipticity of each source is then calculated by adding, to the interpolated shear value, the galaxy's true ellipticity rotated by a random angle. This last steps allows us to include shape noise in a realistic way (under the assumption that the ellipticity of a single object is dominated by noise, and not by the lensing-induced shear).
    \end{enumerate}
    
    Each simulated catalogue is then put through the same power spectrum pipeline used for the real data, including masking, power spectrum estimation and field coaddition. We use the power spectra measured in each simulation to estimate the sample covariance matrix as:
    \begin{equation}\label{eq:covar}
      \mathsf{\Sigma}=\frac{1}{N_s-1}\sum_{i=1}^{N_s}\left(\vec{\bf C}_i-\langle\vec{\bf C}\rangle\right)\,\left(\vec{\bf C}_i-\langle\vec{\bf C}\rangle\right)^T,
    \end{equation}
    where $N_s$ is the number of simulations used to estimate the covariance, $\vec{\bf C}_i$ is the vector of power spectra measured in the $i$-th simulation and $\langle\cdot\rangle$ denotes averaging over simulations. When using this covariance to compute Gaussian likelihoods, we correct the inverse covariance by the multiplicative `Hartlap factor' introduced by \cite{2007A&A...464..399H} to account for the finite number of simulations used. As shown by \cite{2016MNRAS.456L.132S}, this approximation is sufficiently accurate given the large number of simulations used in our analysis. This was explicitly confirmed in \cite{2017MNRAS.465.2033J}.
    
    Note that, as described in the next section, we use a maximum of $N_s=2000$ simulations when analysing the full set of power spectra, but we scale this number down with the size of the final data vector when exploring cases with different numbers of KL modes. This is to showcase the usage of data compression to reduce the number of mock realizations needed to estimate the covariance matrix while retaining the final constraining power. In detail, the number of simulations used for each case is chosen to keep the `Hartlap factor' constant.
    
    We neglect all non-Gaussian contributions to the covariance matrix. The only relevant contribution for weak-lensing measurements, as quantified by \cite{2018JCAP...10..053B} is the so-called super-sample covariance, which becomes less important for low-density samples where the measurements on high-$\ell$s are noise-dominated. This fact, together with the acceptable value of the $\chi^2$ we obtain both for the fiducial and best-fit cosmological parameters (see Section \ref{sec:results}) imply that the covariance matrix estimated with this method is appropriate for our analysis. It is worth noting that our objective is not to provide state-of-the-art constraints on cosmological parameters from these data, but rather to showcase the performance of the KL decomposition in terms of data compression. The resulting $E$-mode power spectrum covariance is shown in Figure \ref{fig:covar_cls}. For a given auto- or cross-spectrum, the covariance is strongly dominated by the diagonal.
    
    In order to compare our results with those of \cite{2017MNRAS.465.2033J}, we follow their estimation of the covariance matrix for the real-space correlation functions as closely as possible. In this case, the covariance matrix is estimated from a large suite of 497 distinct Scinet LIght Cone Simulations \citep{2015MNRAS.450.2857H} based on a WMAP9$+$BAO$+$SN cosmology. Following $2\times2$ sub-division of the simulation boxes, the resulting set of 1988 pseudo-independent mock CFHTLenS shear catalogues correspond to a `Hartlap factor' of 0.86, and have the correct effective source density, shape noise, photometric redshift scatter, small-scale masking, and other properties that match the real survey. Given the different effective areas of the four CFHTLenS regions, the final covariance was obtained through area-weighted averaging of the patch-dependent inverse covariances. We refer the reader to \cite{2017MNRAS.465.2033J} for further details. The number of simulations used for the analysis of different sets of KL modes was scaled with the size of the data vector as described in the case of the Fourier-space measurements.

\section{Results}\label{sec:results}
  \subsection{The \kalo transform}\label{ssec:results.kalo}
    \begin{figure}
      \centering
      \includegraphics[width=0.49\textwidth]{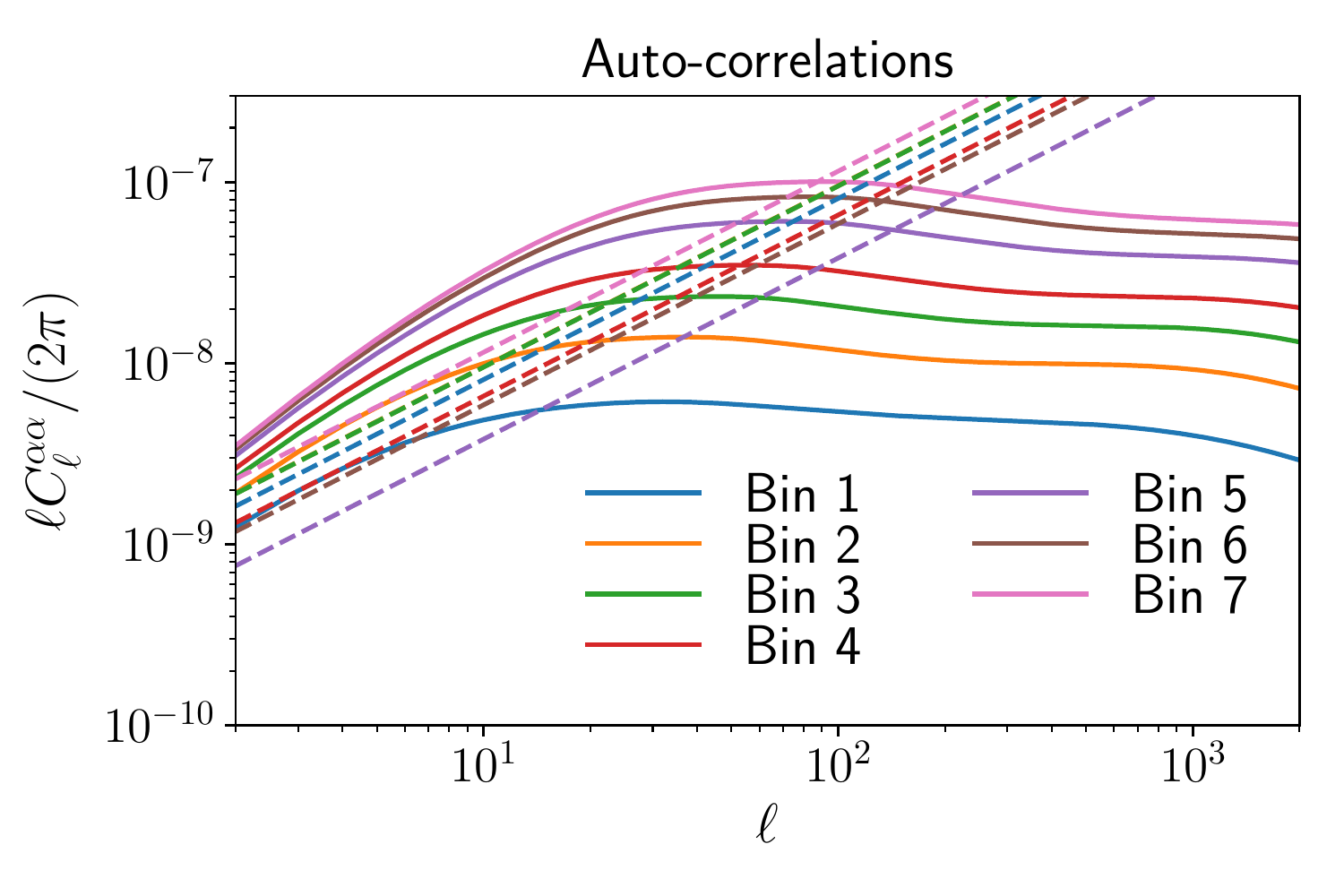}\vspace{-5pt}
      \caption{We show the angular power spectra $C_\ell^{\alpha\alpha}$ and the noise $N_\ell^{\alpha\alpha}$ for each redshift bin auto-correlation as a function of the multipole $\ell$.}\label{fig:cls_diag}
    \end{figure}
    Before presenting the main result of this paper: i.e. the level of data compression that can be achieved through the KL transform, it is useful to review the key steps of the method, as described in Section~\ref{ssec:methods.kl}, and how we apply them in practice to the data. Figure~\ref{fig:cls_diag} shows the diagonal part of the angular $E$-mode power spectra, $C_\ell^{\alpha\alpha}$, as computed by \textsc{CCL}, together with the expected noise $N_\ell^{\alpha\alpha}$ as a function of the multipole $\ell$ for the best fit parameters of \cite{2017MNRAS.465.2033J}, which we use here as our fiducial cosmology. The fact that the noise dominates over the signal for a wide range of scales, and that the off-diagonal signal terms (not shown here) have amplitudes that are similar to the diagonal (given the radially cumulative nature of weak lensing), indicates that different modes are tightly correlated and therefore that the effective number of signal-dominated independent radial modes should be small (i.e. the information carried out by the angular power spectrum is inefficiently spread out all over the redshift bins).

\begin{figure}
      \centering
      \includegraphics[width=0.49\textwidth]{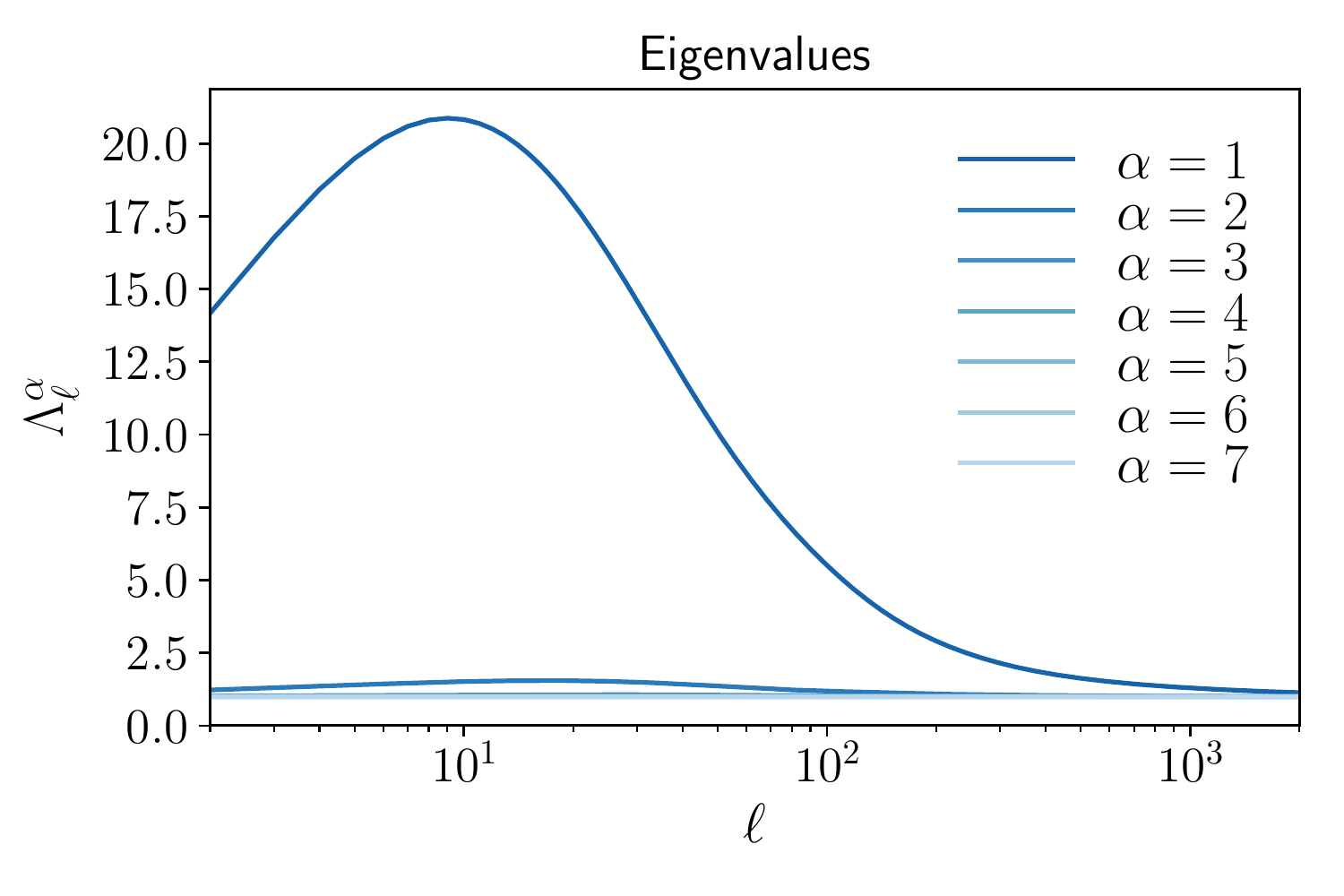}\vspace{-5pt}
      \caption{Theory angular power spectrum for the fiducial cosmological parameters after the KL transform as a function of the multipole $\ell$. The first eigenvalue has the darkest colour and contains most of the information, while the last eigenvalue has the lightest colour and is completely subdominant.}\label{fig:eigvals}
    \end{figure}

    As described in Section~\ref{ssec:methods.kl}, the two advantages of working with KL-transformed power spectra, is that ideally some combination of the redshift bins is constructed to have uncorrelated radial modes and that the information is compressed in the first few modes. This second feature can be seen in Figure \ref{fig:eigvals}, where we show the $\ell$-dependence of the eigenvalues of Eq.~\ref{eq:kl_eigenvalue}, which represent the KL-transformed power spectra (labelled  $\Lambda^{\alpha}_\ell$ here). This quantity measures the amount of signal to noise present in each mode, where the KL-transformed noise power spectrum is simply unity ($\delta_{\alpha\beta}$) for all $\ell$s by construction. It is clear that the majority of the information is concentrated in the first two bins, while the information we lose by considering only the first three bins is almost negligible and the remaining bins are noise dominated.
    \begin{figure}
      \centering
      \includegraphics[width=0.49\textwidth]{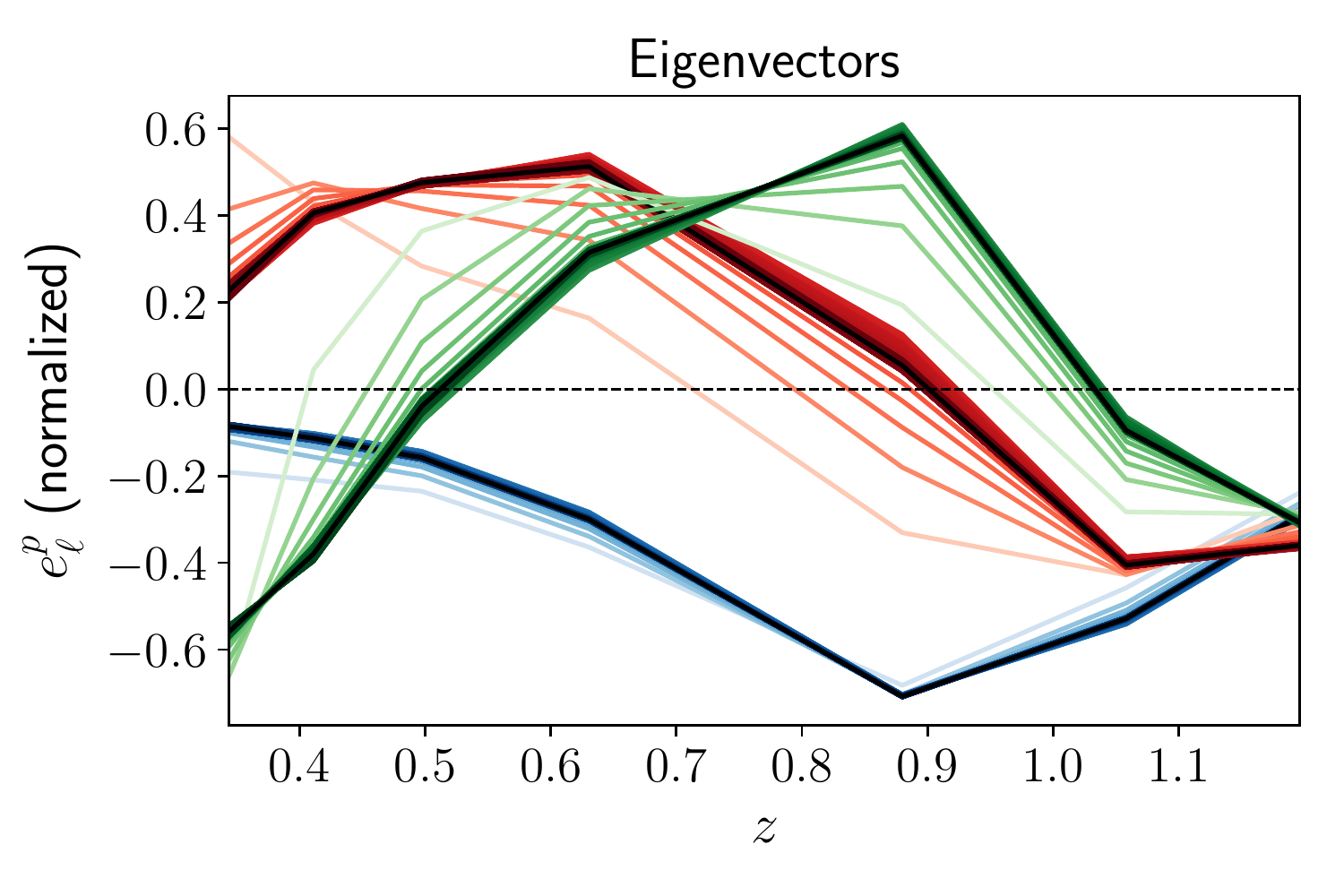}\vspace{-5pt}
      \caption{We show the first three eigenvectors of the KL transform as a function of the redshift bin. Blue lines represent the first eigenvector, red lines the second and green line the third one. Different shades of the same colour indicate the eigenvector for different $\ell$'s (lightest colour refers to $\ell=2$ and increases towards darker shades with $\Delta\ell=10$).}\label{fig:eigvecs}
    \end{figure}

    A further simplification comes by assuming our KL transform to be $\ell$-independent, as proposed in \cite{2018MNRAS.473.4306A}. This is necessary in real space, where $\ell$ multipoles are replaced by angles $\theta$\footnote{Note that the KL transform uses orthogonal eigenfunctions, and this is automatically achieved by decomposing the full sphere into spherical harmonics (or Fourier coefficients in the flat-sky approximation), and therefore the method lives naturally in $\ell$ space.}, but they can be safely used even in Fourier space. This is shown in Figure~\ref{fig:eigvecs}, where we plot the first three principal eigenvectors of Eq.~\ref{eq:kl_eigenvalue} as a function of redshift. Each colour represents a single eigenvector, while different shades refer to different $\ell$s: lighter colours for low-$\ell$s and darker ones for high-$\ell$s. Excluding the first few $\ell$s (where cosmic variance dominates the uncertainties and little information can be recovered), all eigenvectors for the same $\alpha$ clearly have the same shape. Moreover, the blue lines, which represent the first eigenmode and dominate the information content, are the least scattered even at low-$\ell$. Then, an averaged version of the KL transform $e^\alpha_{p,\ell}$ can be written as
    \begin{align}\label{eq:kl_ell_indep}
      &\bar{e}^\alpha_p = \frac{\sum_\ell (2\ell+1)\,e^\alpha_{p,\ell}}{\sum_{\ell} (2\ell+1)}\,,
    \end{align}
    where the $2\ell+1$ factors are included to roughly account for the statistical weight of each harmonic multipole. When performing the real-space analysis, we used this $\ell$-independent KL transform. In Fourier space we tested both options, and we will show that dropping the scale dependence of the eigenvectors makes little or no difference.
    \begin{figure}
      \centering
      \includegraphics[width=0.49\textwidth]{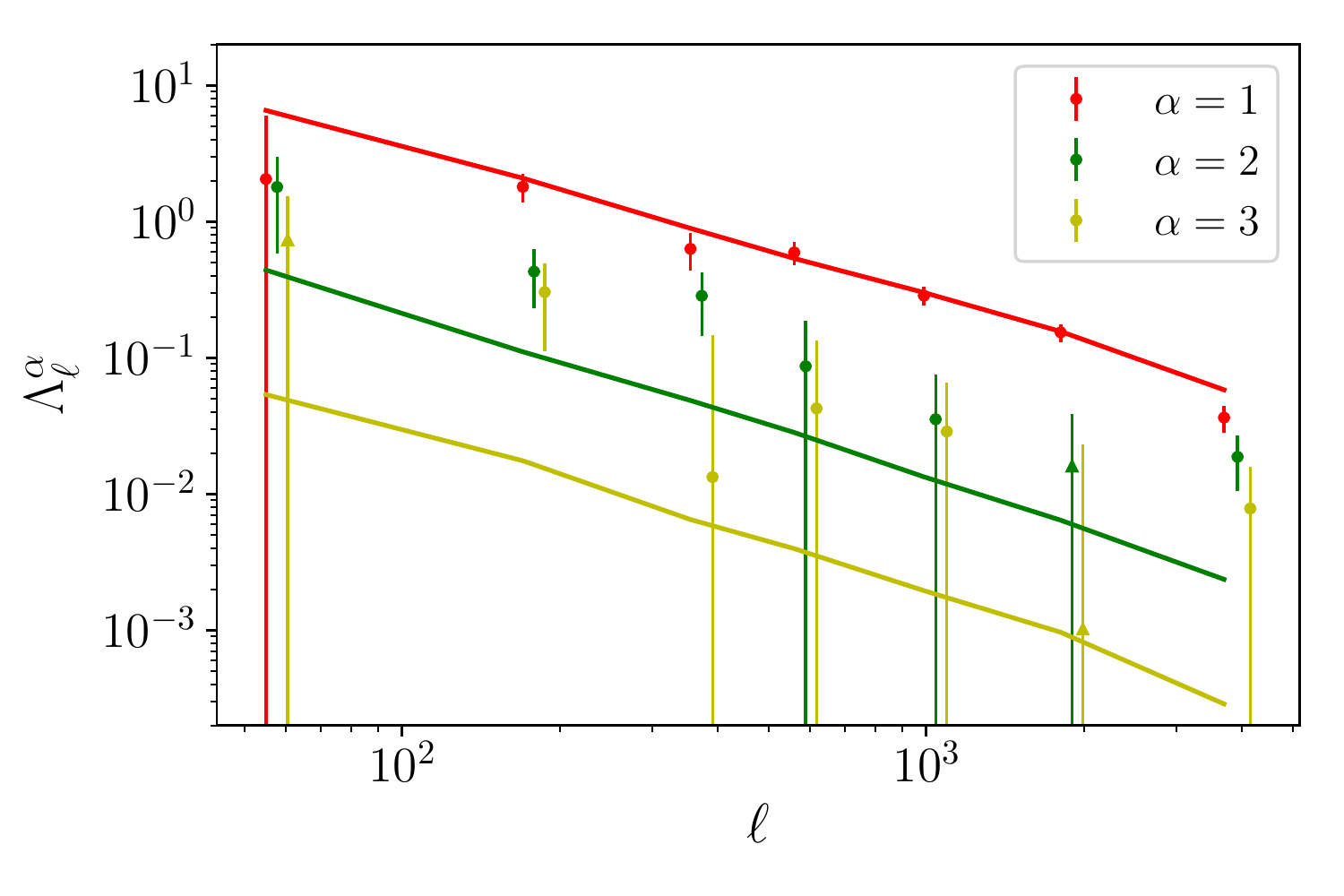}\vspace{-5pt}
      \caption{Measured power spectra for the first three KL eigenvalues as a function of angular scale $\ell$ (points with error bars), together with their theoretical prediction for the fiducial cosmological parameters used in our analysis. Most of the signal is contained in the first eigenvalue, with all other modes being significantly noise-dominated.}\label{fig:eigvals_obs}
    \end{figure}
    
    The measured power spectra for the first three KL eigenmodes using a scale-dependent KL transform are shown in Figure \ref{fig:eigvals_obs}, together with the theoretical prediction corresponding to the fiducial cosmological model used in the analysis. This illustrates how the method concentrates the bulk of the signal-to-noise into the first few modes.
    
    Finally, it is worth emphasizing that, since a critical use of data compression is to reduce the number of simulations needed to estimate the covariance matrix, when analyzing different compression schemes in the next section, we used a covariance matrix computed from a reduced set of simulations, scaled from the full set by the size of the data vector. This allows us to show that achieving a compression factor $f$ will allow us to carry out the analysis using $f$ times fewer simulations.

  \subsection{Observational constraints from the KL decomposition}
    In order to study the efficiency of different data compression schemes, we explore the posterior distribution of cosmological parameters using an MCMC sampler\footnote{In particular we implemented our likelihood in both \textsc{emcee} (\citealt{2013PASP..125..306F}) and \textsc{MontePython} (\citealt{Audren:2012wb}, \citealt{Brinckmann:2018cvx}).}. We assume that the two-point functions (both in real- and Fourier-space) follow a Gaussian likelihood. This is known to be formally incorrect \citep{2018MNRAS.473.2355S,2008PhRvD..77j3013H}, and can give rise to parameter shifts below the $1\sigma$ level. For the purposes of this analysis (i.e. to show the performance of the KL transform in terms of data compression) this effect can be ignored, particularly given the fact that most of the constraining power is obtained from the smaller scales, where the large number of available modes makes the Gaussian approximation acceptable due to the central limit theorem. As in \cite{2017MNRAS.465.2033J}, we use a standard $\Lambda$CDM model and vary five cosmological parameters: the Hubble parameter $h$, the physical baryon density parameter $\Omega_b h^2$, the physical cold dark matter density parameter $\Omega_c h^2$, the primordial comoving curvature power spectrum  amplitude $A_s$ and the spectral index $n_s$. The priors used in this analysis are uniform and reported in Table~\ref{tab:priors}. As investigated in \cite{2017MNRAS.465.2033J}, the impact of priors is non-negligible in determining the allowed parameter space. However, since the scope of this work is just to show the goodness of the KL transform in reducing the data vector without losing any significant information, we choose to keep these priors fixed to those used in previous analyses for all our runs. 
    \begin{table}
      \begin{center}
      \begin{tabular}{c|c}
        \hline
        Parameter & Prior\\
        \hline\hline
        $h$   & $0.61$ -- $0.81$\\
        $\Omega_b h^2$   & $0.013$ -- $0.033$\\
        $\Omega_c h^2$   & $0.001$ -- $0.99$\\
        $\ln\left(10^{10} A_s\right)$   & $2.3$ -- $5.0$\\
        $n_s$   & $0.7$ -- $1.3$\\
        \hline
      \end{tabular}
      \caption{Cosmological parameter priors used in this analysis. All the priors are uniform.}\label{tab:priors}
      \end{center}
    \end{table}

\begin{figure}
      \centering
      \includegraphics[width=0.9\columnwidth]{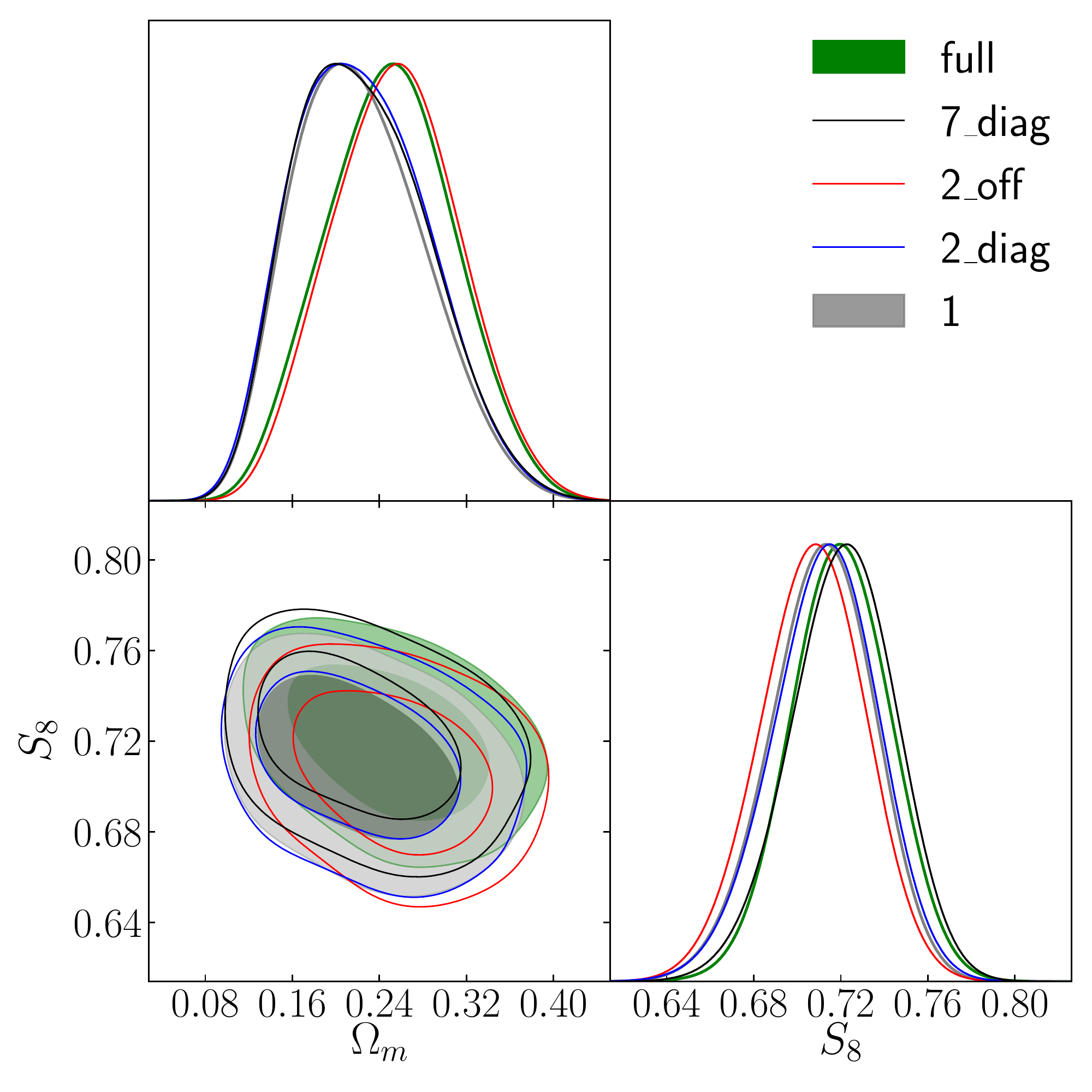}\vspace{-5pt}
      \caption{Marginalized posterior distribution for the parameters $\Omega_m$ and $S_8$. These contours are obtained with a Fourier-space analysis assuming a scale-dependent KL transform. Different colours represent different levels of KL compression: {\tt full} is the full analysis with no KL-compression, {\tt n\_off} is the KL-compression where we consider {\tt n} radial modes and the cross-correlations between them, in {\tt n\_diag} we consider {\tt n} radial modes without the cross-correlations, {\tt 1} shows the maximum compression we can achieve by considering only $1$ mode.}\label{fig:fourier_cont}
    \end{figure}

\begin{figure}
      \centering
      \includegraphics[width=0.9\columnwidth]{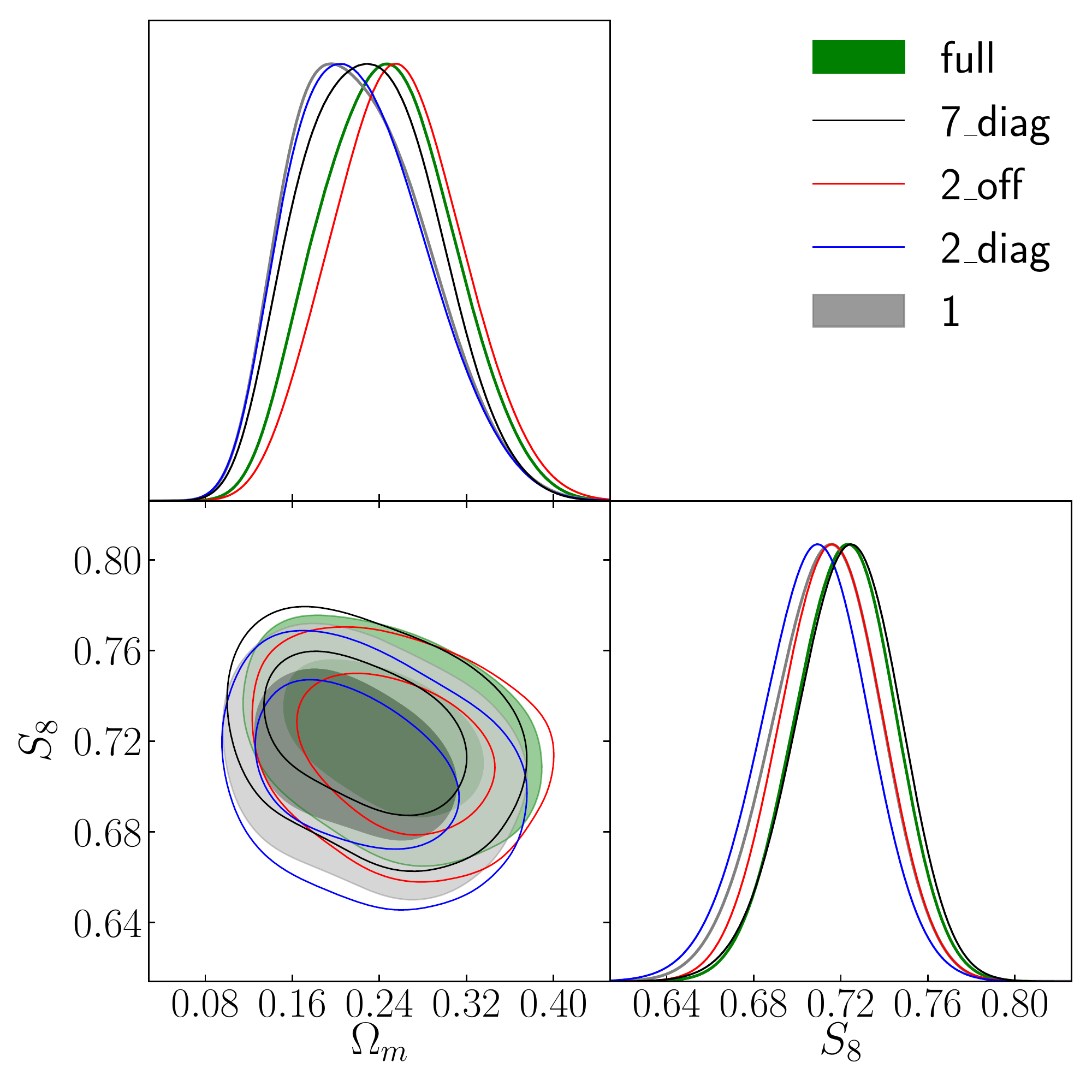}\vspace{-5pt}
      \caption{The same as in Figure~\ref{fig:fourier_cont}, but the KL-transform is assumed to be scale-independent (see Eq.~\ref{eq:kl_ell_indep}).}\label{fig:scale_cont}
    \end{figure}

\begin{figure}
      \centering
      \includegraphics[width=0.9\columnwidth]{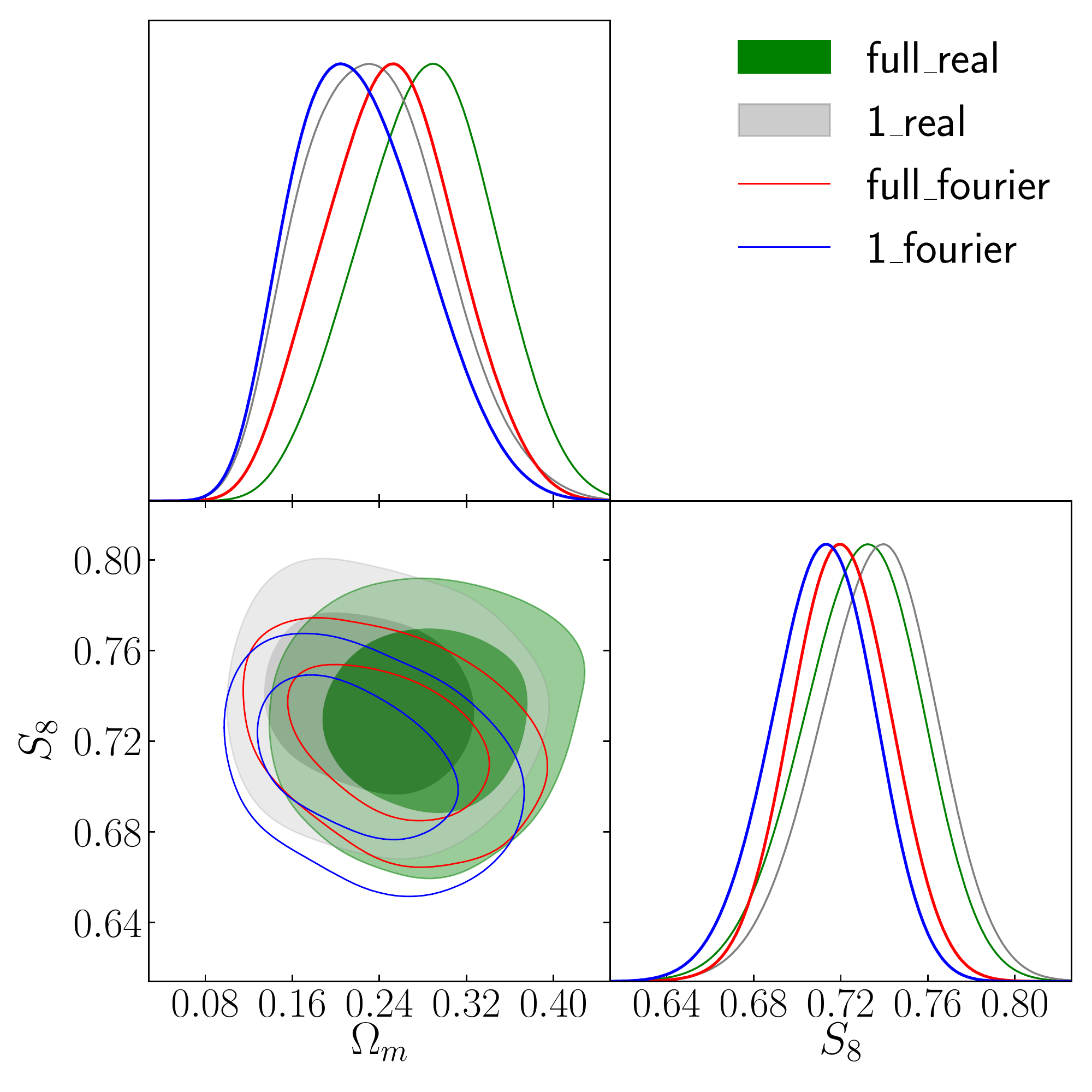}\vspace{-5pt}
      \caption{The same as in Figure~\ref{fig:fourier_cont} but here we compare the contours obtained in the real-space analysis (solid lines and filled contours) with the ones of the Fourier analysis assuming an $\ell$-dependent KL-transform.}\label{fig:real_cont}
    \end{figure}

    In Figure \ref{fig:fourier_cont}, we show the posterior distribution for the parameters $\Omega_m\equiv\Omega_c+\Omega_b$ and $S_8 \equiv\sigma_8\sqrt{\Omega_m/0.3}$ obtained with the analysis in Fourier space and with an $\ell$-dependent KL-transform. The derived quantity $S_8$ is of particular interest in weak lensing analyses since cosmic shear has the most constraining power on this parameter and it has proven to be less dependent on the specific choice of priors. We show a few representative compression schemes: (i) {\tt full} is the case where no KL-compression is used, (ii) in {\tt 7\_diag} we consider all $7$ radial KL modes but only using their auto-correlations, (iii) in {\tt 2\_off} we use the first 2 radial modes but we consider their auto- and cross-correlations, (iv) {\tt 2\_diag} is the same as {\tt 2\_off} but using only auto-correlations, and (v) {\tt 1} is the maximum compression case, where we consider only the first KL mode. Contours for the same parameters are shown in Figures~\ref{fig:scale_cont} and \ref{fig:real_cont} for the Fourier analysis with $\ell$-independent KL-transform and the real-space analysis respectively, where the latter also compares the real-space and Fourier-space results directly. Several conclusions can be drawn from these results:
    \begin{itemize}
      \item We are able to place similar constraints on the most relevant cosmological parameters in terms of final uncertainties in all cases.
      \item We observe small shifts in the maximum-likelihood parameter values for different compression schemes. These are well within the $1\sigma$ level, and are to be expected, since different compression schemes effectively use different sectors of the data vector. A further source of random shifts is the use of different sets of simulations to compute the covariance matrix for the different data compression schemes. We checked using different simulations to compute the covariance matrix that these shifts can be considered as statistical fluctuations.
      \item We obtain qualitatively equivalent results in terms of data compression efficiency using $\ell$-dependent and $\ell$-independent KL modes in a Fourier-space analysis, as well as using a real-space pipeline.
    \end{itemize}
	
    In order to get a quantitative sense of the compression efficiency of the KL transform, in Table~\ref{tab:fom} we show the Figure-of-Merit (FoM) obtained in the various cases, calculated as the inverse of the areas of the 1$\sigma$ contours in the 2D $\Omega_m$-$S_8$ plane. It is possible to notice a substantial difference between the real-space and the Fourier-space analyses. This is to be expected. We believe that most of these differences can be attributed to the different methods used to estimate the covariance matrices in both cases.\footnote{It is interesting to note that both analyses also obtain substantially different values for the best-fit parameter $\chi^2$ with $\chi^2/{\rm ndof}=1.47$ for the real-space analysis and $\chi^2/{\rm ndof}=0.98$ for the Fourier-space one.} In addition, it is never possible to match both sets of scale cuts, giving rise to different levels of constraining power. More interestingly, for the purposes of data compression, for a given analysis choice we observe only insignificant variations in the FoM as we reduce the number of KL modes left in the data vector. This clearly indicates that the information contained in the {\tt full} case is essentially the same as in the most compressed one.
    Also, by comparing the results for the two different Fourier-space analyses, i.e. the $\ell$-dependent (SD) and the $\ell$-independent (SI) KL transform, we can conclude that there is no significant discrepancy between the two if we take into account cross-correlations between different radial modes. When neglecting the cross-correlations of the power spectra, we find a small ($\sim10\%$) degradation in the FoM for the $\ell$-independent case. This indicates that, while considering an $\ell$-dependent KL transform may retain marginally more information especially when we maximally compress our data vector, the loss of information is small. It is interesting to notice that the FoM sometimes increases when we further compress the data vector. By construction, it is impossible to gain constraining power by compressing the data. This indicates that the apparent increasing of the FoM can be interpreted only as a statistical fluctuation.

    In addition, since $S_8$ is the best constrained parameter with current weak lensing surveys, we can estimate the goodness of the KL compression by looking at its 1D posterior distribution. We compared the absolute values of the relative shift in the means ($\left|\Delta S_8/\sigma(S_8)\right|$), and of the relative difference of the standard deviations ($\left|\Delta \sigma(S_8)/\sigma(S_8)\right|$) between the \texttt{full} and the {\tt 1} cases. These two have been chosen as representative of the cases with none and maximum data compression respectively. For the shift in the means we find $\sim0.38$ for both the Fourier-space analyses (SD and SI) and $\sim0.25$ for the real-space analysis. Moreover, the relative difference of the standard deviations is $\sim0.04$ for the Fourier-space SD analysis, $\sim0.08$ for the Fourier-space SI analysis and $\sim0.01$ for the real-space analysis. This indicates not only that the amount of information lost with the KL-transform is practically negligible, but also that the shifts in the posterior distributions are compatible with expected statistical fluctuations caused by ignoring part of the signal.
    
    In the best case, comparing all the data points used in the real-space analysis of \cite{2017MNRAS.465.2033J} with the maximum data compression of the Fourier-space pipeline, we are able to achieve a compression factor of $\simeq 30$, which greatly simplifies the problem of computing the covariance matrix from both mock catalogues and with analytic methods.

    \begin{table}
    \begin{center}
    \begin{tabular}{cc|ccc}
    \hline
    KL & Size & Fourier SI & Fourier SD & Real\\
    \hline\hline
    {\tt full}   & $28\times s_{\ell\theta}$ & $109.7$ & $109.8$ & $83.0$\\
    {\tt 6\_off} & $21\times s_{\ell\theta}$ & $103.0$ & $104.6$ & $82.1$\\
    {\tt 5\_off} & $15\times s_{\ell\theta}$ & $107.7$ & $106.2$ & $85.4$\\
    {\tt 4\_off} & $10\times s_{\ell\theta}$ & $102.7$ & $109.0$ & $83.1$\\
    {\tt 3\_off} & $6\times s_{\ell\theta}$ & $104.2$ & $100.1$ & $85.2$\\
    {\tt 2\_off} & $3\times s_{\ell\theta}$ & $105.8$ & $103.1$ & $89.5$\\
    \hline
    {\tt 7\_diag} & $7\times s_{\ell\theta}$ & $104.5$ & $102.9$ & $81.0$\\
    {\tt 6\_diag} & $6\times s_{\ell\theta}$ & $104.1$ & $114.0$ & $73.0$\\
    {\tt 5\_diag} & $5\times s_{\ell\theta}$ & $105.0$ & $102.3$ & $76.4$\\
    {\tt 4\_diag} & $4\times s_{\ell\theta}$ & $106.6$ & $115.9$ & $72.4$\\
    {\tt 3\_diag} & $3\times s_{\ell\theta}$ &  $99.9$ & $111.8$ & $76.6$\\
    {\tt 2\_diag} & $2\times s_{\ell\theta}$ &  $98.9$ & $101.1$ & $81.7$\\
    {\tt 1}       & $1\times s_{\ell\theta}$ &  $98.8$ & $107.0$ & $83.0$\\
    \hline
    \end{tabular}
    \caption{We show the size of the data vector along with the Figure of Merit (FoM) for each case considered here. The first column indicates the KL-compression used. The second column shows the size of the data vector for each method, where $s_{\ell\theta}=5$ in Fourier space ($5$ $\ell$ bandpowers) and $s_{\ell\theta}=10$ in Real space ($6$ angles $\theta$ for $\xi_+$ and $4$ for $\xi_-$). The other columns show the FoM for each case, calculated as the Area$^{-1}$ of the $1\sigma$ contour in $\Omega_m$-$S_8$. ``Fourier SI'' refers to the analysis in Fourier Space with $\ell$-independent KL-transform, ``Fourier SD'' refers to the $\ell$-dependent KL-transform, while ``Real'' shows the KL-transformed results in real space.}\label{tab:fom}
    \end{center}
    \end{table}

    \subsection{Relaxing the assumptions}\label{sec:relax}
    
    The constraints obtained in the previous section show that even with the maximum KL compression the loss of information is negligible. However, they were obtained using a number of assumptions that can be relaxed in order to test the robustness of our conclusion.
    
    \begin{figure}
      \centering
      \includegraphics[width=0.9\columnwidth]{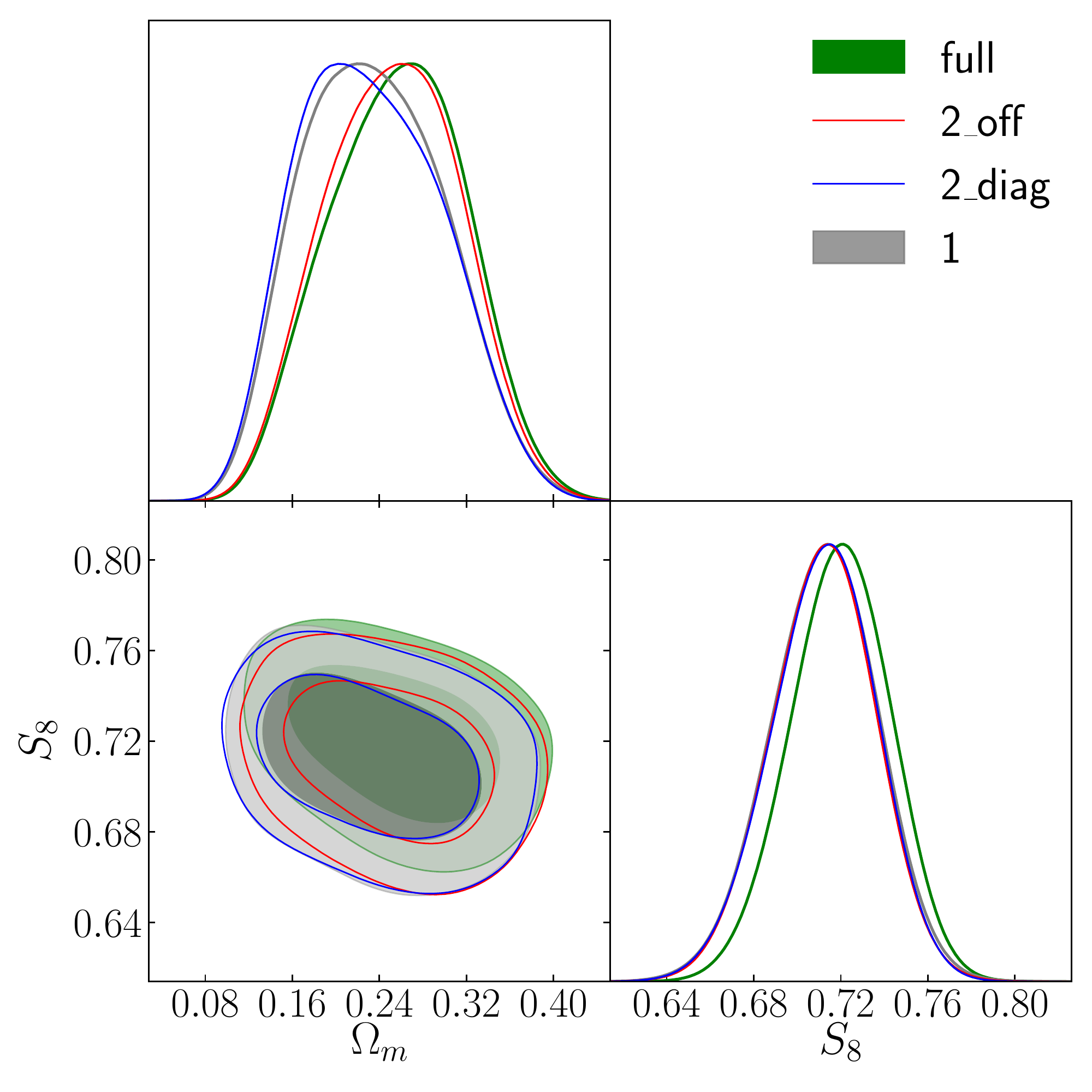}\vspace{-5pt}
      \caption{The same as in Figure~\ref{fig:fourier_cont} but the KL transform has been calculated using a bad fit to the data, i.e.~$\chi^2\sim2\times\chi^2_{\rm best fit}$.}\label{fig:bad_cont}
    \end{figure}
    As the KL transform has to be calculated assuming a fiducial model, the loss of information is minimal if the fiducial model is a good fit to the data. This is the main reason to choose the best fit of \cite{2017MNRAS.465.2033J} as our fiducial model. To explore the impact of the fiducial cosmology on the performance of this method, we have repeated our analysis using a fiducial model that is significantly far away from the best fit. In particular we use a flat $\Lambda$CDM model with $(h,\,\Omega_bh^2,\,\Omega_ch^2,\ln(10^{10}A_s),n_s)=$ $(0.8,0.02,0.16,3.2,0.8)$, which yields a $\chi^2$ for the \cfh data that is twice as large as that of the best fit. The result is shown in Figure \ref{fig:bad_cont}, which shows the posterior distribution for the parameters $\Omega_m$ and $S_8$. We find contours that are very similar to the ones shown in Figure \ref{fig:scale_cont} and, as in the previous section, we find that the shift in the means and the relative difference of the 1-$\sigma$ errors for $S_8$ are $\sim0.32$ and $\sim0.06$ respectively. The performance of the KL transform is therefore only mildly affected by the choice of fiducial model.

    \begin{figure}
      \centering
      \includegraphics[width=0.9\columnwidth]{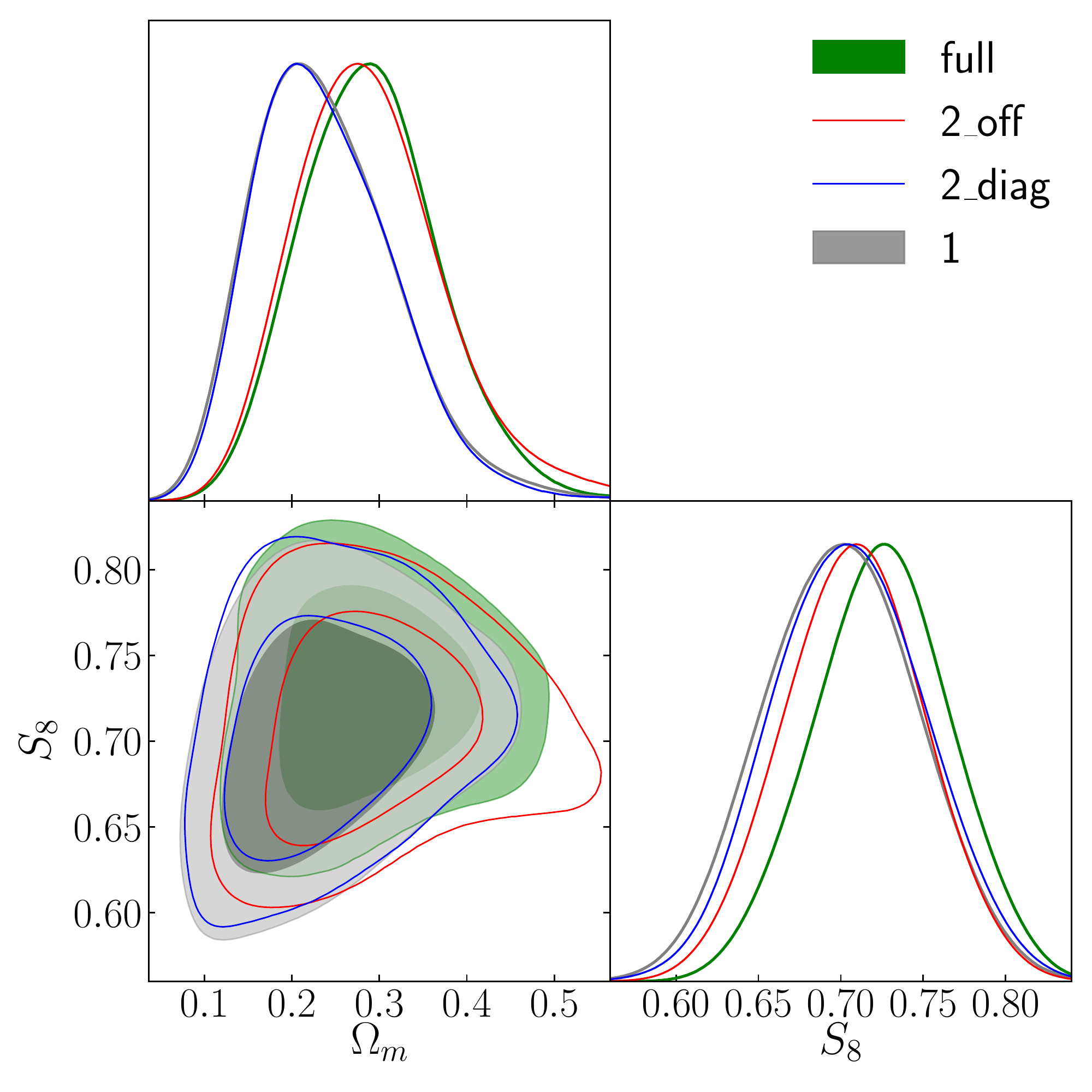}
      \includegraphics[width=0.9\columnwidth]{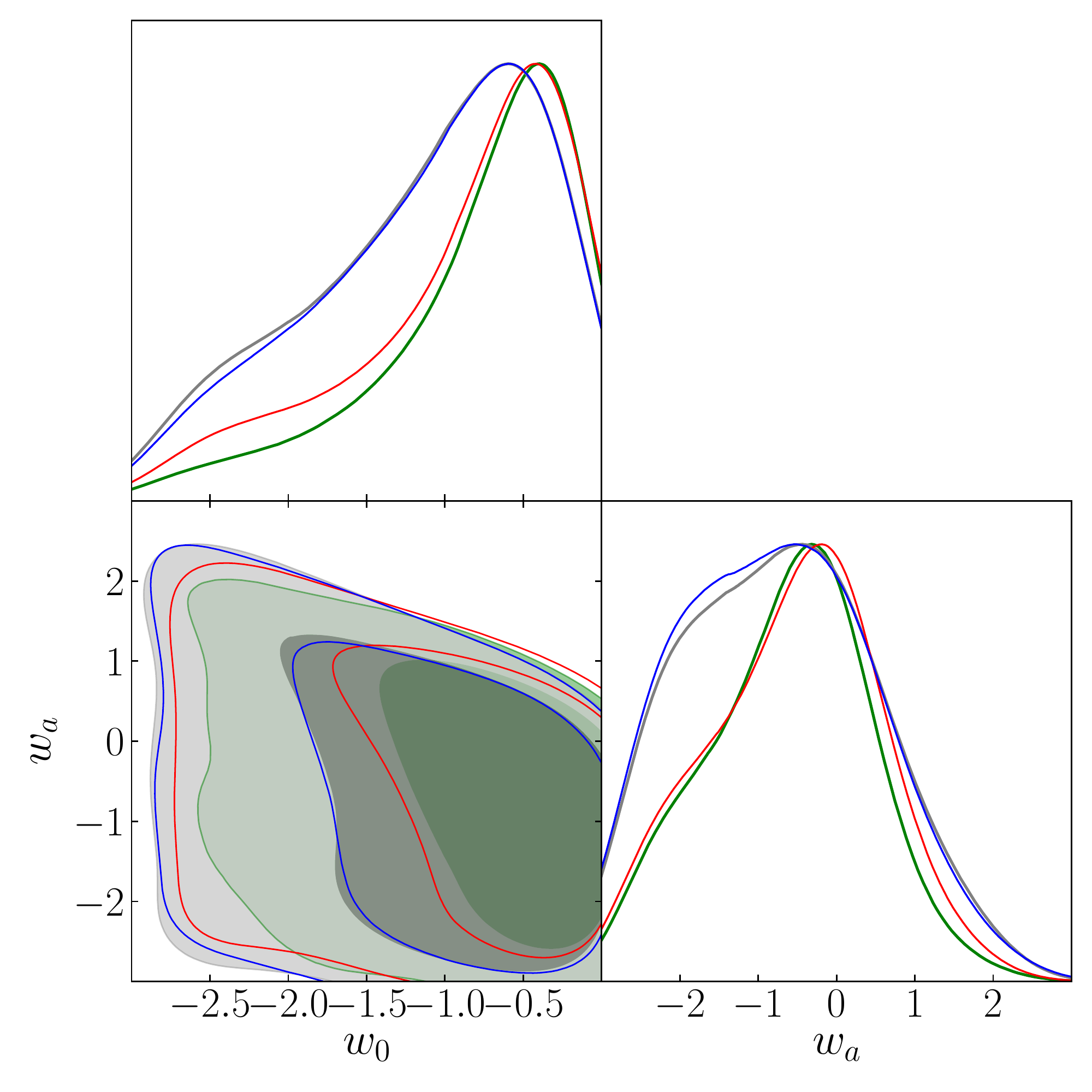}\vspace{-5pt}
      \caption{The same as in Figure~\ref{fig:fourier_cont} but replacing the cosmological constant $\Lambda$ with a time-varying equation of state, $w=w_0+(1-a)w_a$. The top panel shows the constraints in the $\Omega_M$-$S_8$ plane, while the bottom one shows the results for $\{w_0,\,w_a\}$. This extended parameter space evidences the benefits of including a second KL mode in the analysis.}\label{fig:w0wa_cont}
    \end{figure}
    Moreover, the results of our main analysis are in principle only valid for standard flat $\Lambda$CDM models described by the parameters in Table \ref{tab:priors}. An extended parameter space could in principle benefit from including additional KL modes in order to break parameter degeneracies. To explore this possibility, we have repeated our analysis for a ``$w$CDM'' model, in which we parametrize the evolution of the equation of state of dark energy as $w(a)=w_0+(1-a)w_a$ \citep{2001IJMPD..10..213C,2003PhRvL..90i1301L}, where $a$ is the scale factor, and $(w_0,w_a)$ are two new free parameters of the model\footnote{We impose flat priors on these parameters with ranges $w_0\in[-3,0]$ and $w_a\in[-3,3]$, while leaving all other priors fixed to the values in Table~\ref{tab:priors}.}. The results are shown in Figure \ref{fig:w0wa_cont} both for the $\Omega_M$-$S_8$ plane (top panel) and the $w_0$-$w_a$ plane (bottom panel). As expected, the parameter contours for $\Omega_M$-$S_8$ are broader than in the $\Lambda$CDM case. The corresponding parameter shift and degradation in the final error on $S_8$ when comparing the {\tt full} and {\tt 1} cases are $\Delta S_8/\sigma_8=0.58$, $\Delta\sigma(S_8)/\sigma(S_8)=0.12$. These are noticeably larger than in the $\Lambda$CDM case, reflecting the fact that additional information is encoded in the remaining KL modes. This is more evident in the $w_0$-$w_a$ plane, where the uncertainties on both parameters degrade noticeably if only a single mode is used. However, we can see that, by including both the first and second KL modes and considering their cross-correlation (dashed violet contours), we can recover the same constraints found when using the full data vector (which is also true for the $\Omega_M$-$S_8$ plane). Therefore, the compression efficiency degrades in this case from a factor $\sim30$ to a factor $\sim10$, which is still significant.
    
    This result should come as no surprise, since figures \ref{fig:eigvals} and \ref{fig:eigvals_obs} show that, even if the $S/N$ ratio is dominated by the first KL mode, some information can still be extracted from the second one. In the context of the $w$CDM model, this can be interpreted as follows: although the overall amplitude of the matter fluctuations can be well captured by a single, high signal-to-noise measurement of the weak lensing power spectrum, corresponding to the first KL mode, information about their growth as a function of time is vital in order to distinguish between different dark energy models. As shown by the red lines in Fig.~\ref{fig:eigvecs}, the second KL mode weights the high-redshift and low-redshift data differently, and is therefore able to partially recover the growth history. This was also shown in \cite{2018MNRAS.473.4306A}, where it was found that an LSST-like experiment will require at least 3 KL modes in order to recover optimal constraints in an evolving dark energy scenario with massive neutrinos. It is also worth emphasizing the fact that, as shown by this exercise, even though the KL modes are uncorrelated in the fiducial cosmology, additional information can be obtained by exploring the parameter dependence of their cross-correlation.
    
    This result shows that the performance of the KL decomposition depends upon the parameter space on which it is used, and therefore a preliminary forecast is required for any experiment before deciding upon a given compression scheme. Note that this is also a required step before defining the redshift binning to be used in a standard tomographic analysis. It is also worth noting that our particular implementation of the KL transform was aimed to maximize the overall $S/N$ of the data. The method is however more general than that, and can be tuned to optimize the final constraints on any parameter. Therefore, the results shown here could potentially be improved if we fine-tuned our KL decomposition for e.g.~$w_0$ and/or $w_a$.
    
  \subsection{Comparison with other analyses}\label{sec:comparison}
    We compare here the performance of the method described in this paper to other similar analyses carried out on the \cfh data.
    
    The closest study of a radial decomposition of weak lensing data is that of \cite{2014MNRAS.442.1326K}, which uses spherical Bessel functions as a radial basis. This analysis makes use of 50 radial modes, compared to our 1 or 2 KL eigenmodes. The fact that their basis is orthogonal by construction also implies that the cross-correlations between different modes cannot be neglected (see e.g. their Figure 4). One advantage of the modes used by \cite{2014MNRAS.442.1326K} is that, for accurate enough photometric redshifts, each of them can be associated with a physical wavenumber $k$, and therefore the effect of baryons and non-linearities in the matter power spectrum can be kept under control by using a conservative cut on $k$. It is also worth noting that \cite{2014MNRAS.442.1326K} use the map-level modes as their data, instead of summarising them into power spectra. The total number of modes used at the likelihood level is therefore very large, but the use of map-level quantities simplifies the problem of computing 2-point function covariances.
    
    Another interesting analysis is that of \cite{2017MNRAS.464.1676A}, in which the authors make use of a compressed set of Complete Orthogonal Sets of E$/$B-mode Integrals (COSEBIs) calculated from the \cfh data. Their analysis proposes the use of a set of 20 COSEBIs for these data, which corresponds to a data reduction factor of $\sim7$ with respect to their raw initial data vector. Although this is not checked explicitly in this analysis, the resulting compressed COSEBIs should preserve the constraining power on a basic set of 5 cosmological parameters.
    
    Finally, \cite{2015MNRAS.449.1505S} have made use of a KL decomposition to reduce the dimensionality of third-order cosmic shear statistics in \cfh. The compression factor achieved here is modest ($\sim4$), but the use of this type of analyses for these datasets is extremely relevant, given the large number of possible three-point function combinations.

\section{Discussion}\label{sec:discussion}
  The main goal of this paper has been to show the potential of the \kalo transform in reducing the dimensionality of the data vector for cosmic shear measurements. We applied our method to weak lensing data from the \cfh survey. The strong correlation between the weak lensing signal at different redshifts, makes cosmic shear particularly well suited for this type of compression. The fact that other 3D data decomposition frameworks have been applied to the \cfh data in the past additionally allows us to make a comparison between different strategies.
  
  The KL transform is designed to determine the linear combinations of the data that maximize the amount of information about a given parameter. Our main analysis of the angular two-point functions has been performed in Fourier space, both considering an $\ell$-dependent and an $\ell$-independent KL transform that maximizes the $S/N$ of the data. In this context, the resulting set of radial eigenfunctions can be likened to a Fourier decomposition along the redshift coordinate (i.e. the KL transform is a generalization of the standard 3D power spectrum analysis to weak lensing data). For comparison with previous literature, we have also applied our method to real-space correlation functions, and show how the compression works in the different cases. In Fourier space we used the \cfh catalogue to generate masks and maps of the observed area of the sky. We then used these maps to estimate the weak lensing power spectra. For the real-space analysis, we directly used the correlation functions and covariance matrix constructed in \cite{2017MNRAS.465.2033J}. In all cases, theoretical predictions for the power spectra and correlation functions given the cosmological parameters were obtained using \textsc{CCL}. In order to calculate the KL transform, our fiducial point was the best fit of \cite{2017MNRAS.465.2033J}. Choosing cosmological parameters close to the best fit ensures a nearly optimal compression, while random parameters would potentially degrade the efficacy of the method. We then applied the KL transform to our two-point functions and explored the posterior parameter distribution using MCMC methods.

  The main result of this paper is that, both in real and Fourier spaces, the KL compression works exceptionally well. Indeed, it is possible to compress the original data vector, with $280$ and $140$ elements in real and Fourier space respectively, down to a vector with $10$ and $5$ elements without any appreciable loss of information or parameter biases. In other words, we achieve a compression factor of $\sim30$ for CFHTLenS, which simplifies the problem of computing covariance matrices from mock catalogues by the same factor, or by a factor $\sim800$ using analytic methods. This is shown in Table~\ref{tab:fom}, where only insignificant deviations in the FoM are found as we reduce the number of KL modes considered. Additionally, we observe that considering an $\ell$-independent KL transform does not significantly degrade the constraints, which simplifies the application of this method to real-space analyses.
  
  As shown in Section \ref{sec:relax}, almost the same result holds after relaxing some of our assumptions. First, we have shown that the performance of the method is robust against errors in the fiducial cosmology chosen to define the KL eigenmodes. To show this we repeated our analysis for a fiducial cosmology that is substantially disfavoured by the \cfh data, finding no significant difference in the final constraints. We also extended the parameter space by replacing the cosmological constant with a fluid with time dependent equation of state, using the standard $(w_0,w_a)$ parametrization. In this case, we find that, even though the final constraints are affected by the prior on these extended parameters, additional information can indeed be gained by including a second KL mode. This can be expected, given the non-zero signal contained in this mode, and can be interpreted as the second mode being able to recover growth information. The compression factor in this case degrades by a factor $3$ with respect to the $\Lambda$CDM analysis, although this could potentially be further improved by using a KL transform that optimizes the recovery of specific parameters (e.g. $w_0$ and/or $w_a$).

  We must note a number of caveats in our analysis. Most prominently, we have not marginalized over possible photometric redshift systematics, including catastrophic outliers. Standard tomographic approaches have traditionally encapsulated these as additional nuisance parameters per redshift bin (e.g.~overall shifts in the redshift distribution) that are marginalized over when evaluating the likelihood. This is not necessarily appropriate for the KL decomposition used here. The analogous approach would be to marginalize over the redshift distributions associated to each KL mode used ($\tilde{n}_{p}(z)=e_p^\alpha n_\alpha(z)$, using the notation in Section \ref{sec:methods}). Intrinsic alignments \citep{2015PhR...558....1T} is another systematic that may affect the performance of the method, given its non-local nature. Given the low significance of this effect, compared to the expected lensing signal, we expect it to have a negligible impact on the set of optimal eigenmodes, which are defined purely on the basis of signal-to-noise ratio. However, it could degrade significantly the compression efficiency in terms of final parameter constraints. We leave this study for future work. We have also ignored systematics associated to shear calibration, source clustering or baryonic effects on the matter power spectrum. Current and future imaging surveys draw most of their cosmological constraining power from the joint analysis of galaxy clustering and weak lensing (\citealt{vanUitert:2017ieu}, \citealt{Joudaki:2017zdt}, \citealt{2018PhRvD..98d3526A}). It would therefore be interesting to explore the different ways in which the KL transform could be applied to this combined analysis. Although a direct application of the method would yield KL modes that mix clustering and shear, an alternative would be to combine the KL modes of both probes taken individually, in order to separate the different systematic uncertainties that affect them. We leave these investigations for future work.

  Finally, it is worth noting that a number of alternative data compression methods have been proposed in the literature that can achieve significantly larger compression factors (e.g.~\citealt{2001MNRAS.327..849R,2017MNRAS.472.4244H,2018MNRAS.476L..60A,2018MNRAS.477.2874A}), down to a single data point per free parameter of the model. Although some aspects of these methods depend on having a sufficiently large sample of simulations, they are some of the most promising avenues to alleviate the complexity of computing covariance matrices (or even to forgo likelihood computations altogether). One possible advantage of the KL transform as presented here with respect to these methods is the availability of intermediate data products (KL mode maps and power spectra) that can be inspected to identify systematics in the analysis pipeline or the theory model. Data compression techniques of any form will be of tremendous use in future surveys such as LSST or Euclid, simplifying computationally demanding tasks, such as the computation of large covariance matrices.

\section*{Acknowledgements}
We thank Chris Blake for the use of CFHTLenS mock catalogues in the creation of the covariance matrix of the shear correlation functions. We also thank Chris Blake, Pedro Ferreira, Alan Heavens, Catherine Heymans, Lance Miller and Licia Verde for useful discussions. EB and SJ are supported by ERC H2020 693024 GravityLS project, the Beecroft Trust and the Science and Technology Facilities Council (STFC). DA acknowledges support from the Beecroft Trust and from STFC through an Ernest Rutherford Fellowship, grant reference ST/P004474/1. This work is based on observations obtained with MegaPrime/MegaCam, a joint project of CFHT and CEA/DAPNIA, at the Canada-France-Hawaii Telescope (CFHT) which is operated by the National Research Council (NRC) of Canada, the Institut National des Sciences de l'Univers of the Centre National de la Recherche Scientifique (CNRS) of France, and the University of Hawaii. This research used the facilities of the Canadian Astronomy Data Centre operated by the National Research Council of Canada with the support of the Canadian Space Agency. CFHTLenS data processing was made possible thanks to significant computing support from the NSERC Research Tools and Instruments grant program.
  
\setlength{\bibhang}{2.0em}
\setlength\labelwidth{0.0em}
\bibliography{paper}

\begin{thebibliography}{66}
\expandafter\ifx\csname natexlab\endcsname\relax\def\natexlab#1{#1}\fi
\expandafter\ifx\csname bibnamefont\endcsname\relax
  \def\bibnamefont#1{#1}\fi
\expandafter\ifx\csname bibfnamefont\endcsname\relax
  \def\bibfnamefont#1{#1}\fi
\expandafter\ifx\csname citenamefont\endcsname\relax
  \def\citenamefont#1{#1}\fi
\expandafter\ifx\csname url\endcsname\relax
  \def\url#1{\texttt{#1}}\fi
\expandafter\ifx\csname urlprefix\endcsname\relax\def\urlprefix{URL }\fi
\providecommand{\bibinfo}[2]{#2}
\providecommand{\eprint}[2][]{\url{#2}}

\bibitem[{\citenamefont{{Hu}}(1999)}]{1999ApJ...522L..21H}
\bibinfo{author}{\bibfnamefont{W.}~\bibnamefont{{Hu}}}, \bibinfo{journal}{\apj}
  \textbf{\bibinfo{volume}{522}}, \bibinfo{pages}{L21} (\bibinfo{year}{1999}),
  \eprint{astro-ph/9904153}.

\bibitem[{\citenamefont{{Amara} and
  {R{\'e}fr{\'e}gier}}(2007)}]{2007MNRAS.381.1018A}
\bibinfo{author}{\bibfnamefont{A.}~\bibnamefont{{Amara}}} \bibnamefont{and}
  \bibinfo{author}{\bibfnamefont{A.}~\bibnamefont{{R{\'e}fr{\'e}gier}}},
  \bibinfo{journal}{\mnras} \textbf{\bibinfo{volume}{381}},
  \bibinfo{pages}{1018} (\bibinfo{year}{2007}), \eprint{astro-ph/0610127}.

\bibitem[{\citenamefont{{Linder}}(2003)}]{2003PhRvL..90i1301L}
\bibinfo{author}{\bibfnamefont{E.~V.} \bibnamefont{{Linder}}},
  \bibinfo{journal}{\prl} \textbf{\bibinfo{volume}{90}}, \bibinfo{eid}{091301}
  (\bibinfo{year}{2003}), \eprint{astro-ph/0208512}.

\bibitem[{\citenamefont{{Dalal} et~al.}(2008)\citenamefont{{Dalal}, {Dor{\'e}},
  {Huterer}, and {Shirokov}}}]{2008PhRvD..77l3514D}
\bibinfo{author}{\bibfnamefont{N.}~\bibnamefont{{Dalal}}},
  \bibinfo{author}{\bibfnamefont{O.}~\bibnamefont{{Dor{\'e}}}},
  \bibinfo{author}{\bibfnamefont{D.}~\bibnamefont{{Huterer}}},
  \bibnamefont{and}
  \bibinfo{author}{\bibfnamefont{A.}~\bibnamefont{{Shirokov}}},
  \bibinfo{journal}{\prd} \textbf{\bibinfo{volume}{77}}, \bibinfo{eid}{123514}
  (\bibinfo{year}{2008}), \eprint{0710.4560}.

\bibitem[{\citenamefont{{Alonso} and {Ferreira}}(2015)}]{2015PhRvD..92f3525A}
\bibinfo{author}{\bibfnamefont{D.}~\bibnamefont{{Alonso}}} \bibnamefont{and}
  \bibinfo{author}{\bibfnamefont{P.~G.} \bibnamefont{{Ferreira}}},
  \bibinfo{journal}{\prd} \textbf{\bibinfo{volume}{92}}, \bibinfo{eid}{063525}
  (\bibinfo{year}{2015}), \eprint{1507.03550}.

\bibitem[{\citenamefont{{Hu} et~al.}(1998)\citenamefont{{Hu}, {Eisenstein}, and
  {Tegmark}}}]{1998PhRvL..80.5255H}
\bibinfo{author}{\bibfnamefont{W.}~\bibnamefont{{Hu}}},
  \bibinfo{author}{\bibfnamefont{D.~J.} \bibnamefont{{Eisenstein}}},
  \bibnamefont{and}
  \bibinfo{author}{\bibfnamefont{M.}~\bibnamefont{{Tegmark}}},
  \bibinfo{journal}{\prl} \textbf{\bibinfo{volume}{80}}, \bibinfo{pages}{5255}
  (\bibinfo{year}{1998}), \eprint{astro-ph/9712057}.

\bibitem[{\citenamefont{{DES Collaboration} et~al.}(2017)\citenamefont{{DES
  Collaboration}, {Abbott}, {Abdalla}, {Alarcon}, {Aleksi{\'c}}, {Allam},
  {Allen}, {Amara}, {Annis}, {Asorey} et~al.}}]{2017arXiv170801530D}
\bibinfo{author}{\bibnamefont{{DES Collaboration}}},
  \bibinfo{author}{\bibfnamefont{T.~M.~C.} \bibnamefont{{Abbott}}},
  \bibinfo{author}{\bibfnamefont{F.~B.} \bibnamefont{{Abdalla}}},
  \bibinfo{author}{\bibfnamefont{A.}~\bibnamefont{{Alarcon}}},
  \bibinfo{author}{\bibfnamefont{J.}~\bibnamefont{{Aleksi{\'c}}}},
  \bibinfo{author}{\bibfnamefont{S.}~\bibnamefont{{Allam}}},
  \bibinfo{author}{\bibfnamefont{S.}~\bibnamefont{{Allen}}},
  \bibinfo{author}{\bibfnamefont{A.}~\bibnamefont{{Amara}}},
  \bibinfo{author}{\bibfnamefont{J.}~\bibnamefont{{Annis}}},
  \bibinfo{author}{\bibfnamefont{J.}~\bibnamefont{{Asorey}}},
  \bibnamefont{et~al.}, \bibinfo{journal}{arXiv e-prints}
  \bibinfo{eid}{arXiv:1708.01530} (\bibinfo{year}{2017}), \eprint{1708.01530}.

\bibitem[{\citenamefont{{Hildebrandt} et~al.}(2018)\citenamefont{{Hildebrandt},
  {K{\"o}hlinger}, {van den Busch}, {Joachimi}, {Heymans}, {Kannawadi},
  {Wright}, {Asgari}, {Blake}, {Hoekstra} et~al.}}]{2018arXiv181206076H}
\bibinfo{author}{\bibfnamefont{H.}~\bibnamefont{{Hildebrandt}}},
  \bibinfo{author}{\bibfnamefont{F.}~\bibnamefont{{K{\"o}hlinger}}},
  \bibinfo{author}{\bibfnamefont{J.~L.} \bibnamefont{{van den Busch}}},
  \bibinfo{author}{\bibfnamefont{B.}~\bibnamefont{{Joachimi}}},
  \bibinfo{author}{\bibfnamefont{C.}~\bibnamefont{{Heymans}}},
  \bibinfo{author}{\bibfnamefont{A.}~\bibnamefont{{Kannawadi}}},
  \bibinfo{author}{\bibfnamefont{A.~H.} \bibnamefont{{Wright}}},
  \bibinfo{author}{\bibfnamefont{M.}~\bibnamefont{{Asgari}}},
  \bibinfo{author}{\bibfnamefont{C.}~\bibnamefont{{Blake}}},
  \bibinfo{author}{\bibfnamefont{H.}~\bibnamefont{{Hoekstra}}},
  \bibnamefont{et~al.}, \bibinfo{journal}{arXiv e-prints}
  (\bibinfo{year}{2018}), \eprint{1812.06076}.

\bibitem[{\citenamefont{{Hikage} et~al.}(2018)\citenamefont{{Hikage}, {Oguri},
  {Hamana}, {More}, {Mandelbaum}, {Takada}, {K{\"o}hlinger}, {Miyatake},
  {Nishizawa}, {Aihara} et~al.}}]{2018arXiv180909148H}
\bibinfo{author}{\bibfnamefont{C.}~\bibnamefont{{Hikage}}},
  \bibinfo{author}{\bibfnamefont{M.}~\bibnamefont{{Oguri}}},
  \bibinfo{author}{\bibfnamefont{T.}~\bibnamefont{{Hamana}}},
  \bibinfo{author}{\bibfnamefont{S.}~\bibnamefont{{More}}},
  \bibinfo{author}{\bibfnamefont{R.}~\bibnamefont{{Mandelbaum}}},
  \bibinfo{author}{\bibfnamefont{M.}~\bibnamefont{{Takada}}},
  \bibinfo{author}{\bibfnamefont{F.}~\bibnamefont{{K{\"o}hlinger}}},
  \bibinfo{author}{\bibfnamefont{H.}~\bibnamefont{{Miyatake}}},
  \bibinfo{author}{\bibfnamefont{A.~J.} \bibnamefont{{Nishizawa}}},
  \bibinfo{author}{\bibfnamefont{H.}~\bibnamefont{{Aihara}}},
  \bibnamefont{et~al.}, \bibinfo{journal}{arXiv e-prints}
  \bibinfo{eid}{arXiv:1809.09148} (\bibinfo{year}{2018}), \eprint{1809.09148}.

\bibitem[{\citenamefont{{Laureijs} et~al.}(2011)\citenamefont{{Laureijs},
  {Amiaux}, {Arduini}, {Augu{\`e}res}, {Brinchmann}, {Cole}, {Cropper},
  {Dabin}, {Duvet}, {Ealet} et~al.}}]{2011arXiv1110.3193L}
\bibinfo{author}{\bibfnamefont{R.}~\bibnamefont{{Laureijs}}},
  \bibinfo{author}{\bibfnamefont{J.}~\bibnamefont{{Amiaux}}},
  \bibinfo{author}{\bibfnamefont{S.}~\bibnamefont{{Arduini}}},
  \bibinfo{author}{\bibfnamefont{J.~L.} \bibnamefont{{Augu{\`e}res}}},
  \bibinfo{author}{\bibfnamefont{J.}~\bibnamefont{{Brinchmann}}},
  \bibinfo{author}{\bibfnamefont{R.}~\bibnamefont{{Cole}}},
  \bibinfo{author}{\bibfnamefont{M.}~\bibnamefont{{Cropper}}},
  \bibinfo{author}{\bibfnamefont{C.}~\bibnamefont{{Dabin}}},
  \bibinfo{author}{\bibfnamefont{L.}~\bibnamefont{{Duvet}}},
  \bibinfo{author}{\bibfnamefont{A.}~\bibnamefont{{Ealet}}},
  \bibnamefont{et~al.}, \bibinfo{journal}{arXiv e-prints}
  \bibinfo{eid}{arXiv:1110.3193} (\bibinfo{year}{2011}), \eprint{1110.3193}.

\bibitem[{\citenamefont{{LSST Science Collaboration}
  et~al.}(2009)\citenamefont{{LSST Science Collaboration}, {Abell}, {Allison},
  {Anderson}, {Andrew}, {Angel}, {Armus}, {Arnett}, {Asztalos}, {Axelrod}
  et~al.}}]{2009arXiv0912.0201L}
\bibinfo{author}{\bibnamefont{{LSST Science Collaboration}}},
  \bibinfo{author}{\bibfnamefont{P.~A.} \bibnamefont{{Abell}}},
  \bibinfo{author}{\bibfnamefont{J.}~\bibnamefont{{Allison}}},
  \bibinfo{author}{\bibfnamefont{S.~F.} \bibnamefont{{Anderson}}},
  \bibinfo{author}{\bibfnamefont{J.~R.} \bibnamefont{{Andrew}}},
  \bibinfo{author}{\bibfnamefont{J.~R.~P.} \bibnamefont{{Angel}}},
  \bibinfo{author}{\bibfnamefont{L.}~\bibnamefont{{Armus}}},
  \bibinfo{author}{\bibfnamefont{D.}~\bibnamefont{{Arnett}}},
  \bibinfo{author}{\bibfnamefont{S.~J.} \bibnamefont{{Asztalos}}},
  \bibinfo{author}{\bibfnamefont{T.~S.} \bibnamefont{{Axelrod}}},
  \bibnamefont{et~al.}, \bibinfo{journal}{arXiv e-prints}
  \bibinfo{eid}{arXiv:0912.0201} (\bibinfo{year}{2009}), \eprint{0912.0201}.

\bibitem[{\citenamefont{{Newman} et~al.}(2015)\citenamefont{{Newman}, {Abate},
  {Abdalla}, {Allam}, {Allen}, {Ansari}, {Bailey}, {Barkhouse}, {Beers},
  {Blanton} et~al.}}]{2015APh....63...81N}
\bibinfo{author}{\bibfnamefont{J.~A.} \bibnamefont{{Newman}}},
  \bibinfo{author}{\bibfnamefont{A.}~\bibnamefont{{Abate}}},
  \bibinfo{author}{\bibfnamefont{F.~B.} \bibnamefont{{Abdalla}}},
  \bibinfo{author}{\bibfnamefont{S.}~\bibnamefont{{Allam}}},
  \bibinfo{author}{\bibfnamefont{S.~W.} \bibnamefont{{Allen}}},
  \bibinfo{author}{\bibfnamefont{R.}~\bibnamefont{{Ansari}}},
  \bibinfo{author}{\bibfnamefont{S.}~\bibnamefont{{Bailey}}},
  \bibinfo{author}{\bibfnamefont{W.~A.} \bibnamefont{{Barkhouse}}},
  \bibinfo{author}{\bibfnamefont{T.~C.} \bibnamefont{{Beers}}},
  \bibinfo{author}{\bibfnamefont{M.~R.} \bibnamefont{{Blanton}}},
  \bibnamefont{et~al.}, \bibinfo{journal}{Astroparticle Physics}
  \textbf{\bibinfo{volume}{63}}, \bibinfo{pages}{81} (\bibinfo{year}{2015}),
  \eprint{1309.5384}.

\bibitem[{\citenamefont{{Hildebrandt} et~al.}(2017)\citenamefont{{Hildebrandt},
  {Viola}, {Heymans}, {Joudaki}, {Kuijken}, {Blake}, {Erben}, {Joachimi},
  {Klaes}, {Miller} et~al.}}]{2017MNRAS.465.1454H}
\bibinfo{author}{\bibfnamefont{H.}~\bibnamefont{{Hildebrandt}}},
  \bibinfo{author}{\bibfnamefont{M.}~\bibnamefont{{Viola}}},
  \bibinfo{author}{\bibfnamefont{C.}~\bibnamefont{{Heymans}}},
  \bibinfo{author}{\bibfnamefont{S.}~\bibnamefont{{Joudaki}}},
  \bibinfo{author}{\bibfnamefont{K.}~\bibnamefont{{Kuijken}}},
  \bibinfo{author}{\bibfnamefont{C.}~\bibnamefont{{Blake}}},
  \bibinfo{author}{\bibfnamefont{T.}~\bibnamefont{{Erben}}},
  \bibinfo{author}{\bibfnamefont{B.}~\bibnamefont{{Joachimi}}},
  \bibinfo{author}{\bibfnamefont{D.}~\bibnamefont{{Klaes}}},
  \bibinfo{author}{\bibfnamefont{L.}~\bibnamefont{{Miller}}},
  \bibnamefont{et~al.}, \bibinfo{journal}{\mnras}
  \textbf{\bibinfo{volume}{465}}, \bibinfo{pages}{1454} (\bibinfo{year}{2017}),
  \eprint{1606.05338}.

\bibitem[{\citenamefont{{Alonso}}(2018)}]{2018MNRAS.473.4306A}
\bibinfo{author}{\bibfnamefont{D.}~\bibnamefont{{Alonso}}},
  \bibinfo{journal}{\mnras} \textbf{\bibinfo{volume}{473}},
  \bibinfo{pages}{4306} (\bibinfo{year}{2018}), \eprint{1707.08950}.

\bibitem[{\citenamefont{{Heavens}}(2003)}]{2003MNRAS.343.1327H}
\bibinfo{author}{\bibfnamefont{A.}~\bibnamefont{{Heavens}}},
  \bibinfo{journal}{\mnras} \textbf{\bibinfo{volume}{343}},
  \bibinfo{pages}{1327} (\bibinfo{year}{2003}), \eprint{astro-ph/0304151}.

\bibitem[{\citenamefont{{Kitching} et~al.}(2007)\citenamefont{{Kitching},
  {Heavens}, {Taylor}, {Brown}, {Meisenheimer}, {Wolf}, {Gray}, and
  {Bacon}}}]{2007MNRAS.376..771K}
\bibinfo{author}{\bibfnamefont{T.~D.} \bibnamefont{{Kitching}}},
  \bibinfo{author}{\bibfnamefont{A.~F.} \bibnamefont{{Heavens}}},
  \bibinfo{author}{\bibfnamefont{A.~N.} \bibnamefont{{Taylor}}},
  \bibinfo{author}{\bibfnamefont{M.~L.} \bibnamefont{{Brown}}},
  \bibinfo{author}{\bibfnamefont{K.}~\bibnamefont{{Meisenheimer}}},
  \bibinfo{author}{\bibfnamefont{C.}~\bibnamefont{{Wolf}}},
  \bibinfo{author}{\bibfnamefont{M.~E.} \bibnamefont{{Gray}}},
  \bibnamefont{and} \bibinfo{author}{\bibfnamefont{D.~J.}
  \bibnamefont{{Bacon}}}, \bibinfo{journal}{\mnras}
  \textbf{\bibinfo{volume}{376}}, \bibinfo{pages}{771} (\bibinfo{year}{2007}),
  \eprint{astro-ph/0610284}.

\bibitem[{\citenamefont{{McEwen} and {Leistedt}}(2013)}]{2013arXiv1307.1307M}
\bibinfo{author}{\bibfnamefont{J.~D.} \bibnamefont{{McEwen}}} \bibnamefont{and}
  \bibinfo{author}{\bibfnamefont{B.}~\bibnamefont{{Leistedt}}},
  \bibinfo{journal}{arXiv e-prints} \bibinfo{eid}{arXiv:1307.1307}
  (\bibinfo{year}{2013}), \eprint{1307.1307}.

\bibitem[{\citenamefont{{Khalid} et~al.}(2014)\citenamefont{{Khalid},
  {Kennedy}, and {McEwen}}}]{2014arXiv1403.5553K}
\bibinfo{author}{\bibfnamefont{Z.}~\bibnamefont{{Khalid}}},
  \bibinfo{author}{\bibfnamefont{R.~A.} \bibnamefont{{Kennedy}}},
  \bibnamefont{and} \bibinfo{author}{\bibfnamefont{J.~D.}
  \bibnamefont{{McEwen}}}, \bibinfo{journal}{arXiv e-prints}
  \bibinfo{eid}{arXiv:1403.5553} (\bibinfo{year}{2014}), \eprint{1403.5553}.

\bibitem[{\citenamefont{{Leistedt} et~al.}(2015)\citenamefont{{Leistedt},
  {McEwen}, {Kitching}, and {Peiris}}}]{2015PhRvD..92l3010L}
\bibinfo{author}{\bibfnamefont{B.}~\bibnamefont{{Leistedt}}},
  \bibinfo{author}{\bibfnamefont{J.~D.} \bibnamefont{{McEwen}}},
  \bibinfo{author}{\bibfnamefont{T.~D.} \bibnamefont{{Kitching}}},
  \bibnamefont{and} \bibinfo{author}{\bibfnamefont{H.~V.}
  \bibnamefont{{Peiris}}}, \bibinfo{journal}{\prd}
  \textbf{\bibinfo{volume}{92}}, \bibinfo{eid}{123010} (\bibinfo{year}{2015}),
  \eprint{1509.06750}.

\bibitem[{\citenamefont{{Asgari} and {Schneider}}(2015)}]{2015A&A...578A..50A}
\bibinfo{author}{\bibfnamefont{M.}~\bibnamefont{{Asgari}}} \bibnamefont{and}
  \bibinfo{author}{\bibfnamefont{P.}~\bibnamefont{{Schneider}}},
  \bibinfo{journal}{\aap} \textbf{\bibinfo{volume}{578}}, \bibinfo{eid}{A50}
  (\bibinfo{year}{2015}), \eprint{1409.0863}.

\bibitem[{\citenamefont{{Dodelson} and
  {Schneider}}(2013)}]{2013PhRvD..88f3537D}
\bibinfo{author}{\bibfnamefont{S.}~\bibnamefont{{Dodelson}}} \bibnamefont{and}
  \bibinfo{author}{\bibfnamefont{M.~D.} \bibnamefont{{Schneider}}},
  \bibinfo{journal}{\prd} \textbf{\bibinfo{volume}{88}}, \bibinfo{eid}{063537}
  (\bibinfo{year}{2013}), \eprint{1304.2593}.

\bibitem[{\citenamefont{{Petri} et~al.}(2016)\citenamefont{{Petri}, {Haiman},
  and {May}}}]{2016PhRvD..93f3524P}
\bibinfo{author}{\bibfnamefont{A.}~\bibnamefont{{Petri}}},
  \bibinfo{author}{\bibfnamefont{Z.}~\bibnamefont{{Haiman}}}, \bibnamefont{and}
  \bibinfo{author}{\bibfnamefont{M.}~\bibnamefont{{May}}},
  \bibinfo{journal}{\prd} \textbf{\bibinfo{volume}{93}}, \bibinfo{eid}{063524}
  (\bibinfo{year}{2016}), \eprint{1601.06792}.

\bibitem[{\citenamefont{{Krause} and {Eifler}}(2017)}]{2017MNRAS.470.2100K}
\bibinfo{author}{\bibfnamefont{E.}~\bibnamefont{{Krause}}} \bibnamefont{and}
  \bibinfo{author}{\bibfnamefont{T.}~\bibnamefont{{Eifler}}},
  \bibinfo{journal}{\mnras} \textbf{\bibinfo{volume}{470}},
  \bibinfo{pages}{2100} (\bibinfo{year}{2017}), \eprint{1601.05779}.

\bibitem[{\citenamefont{Garc{\'{\i}}a-Garc{\'{\i}}a
  et~al.}(2019)\citenamefont{Garc{\'{\i}}a-Garc{\'{\i}}a, Alonso, and
  Bellini}}]{2019arXiv190611765G}
\bibinfo{author}{\bibfnamefont{C.}~\bibnamefont{Garc{\'{\i}}a-Garc{\'{\i}}a}},
  \bibinfo{author}{\bibfnamefont{D.}~\bibnamefont{Alonso}}, \bibnamefont{and}
  \bibinfo{author}{\bibfnamefont{E.}~\bibnamefont{Bellini}},
  \bibinfo{journal}{Journal of Cosmology and Astroparticle Physics}
  \textbf{\bibinfo{volume}{2019}}, \bibinfo{pages}{043} (\bibinfo{year}{2019}).

\bibitem[{\citenamefont{{Heavens} et~al.}(2017)\citenamefont{{Heavens},
  {Sellentin}, {de Mijolla}, and {Vianello}}}]{2017MNRAS.472.4244H}
\bibinfo{author}{\bibfnamefont{A.~F.} \bibnamefont{{Heavens}}},
  \bibinfo{author}{\bibfnamefont{E.}~\bibnamefont{{Sellentin}}},
  \bibinfo{author}{\bibfnamefont{D.}~\bibnamefont{{de Mijolla}}},
  \bibnamefont{and}
  \bibinfo{author}{\bibfnamefont{A.}~\bibnamefont{{Vianello}}},
  \bibinfo{journal}{\mnras} \textbf{\bibinfo{volume}{472}},
  \bibinfo{pages}{4244} (\bibinfo{year}{2017}), \eprint{1707.06529}.

\bibitem[{\citenamefont{{Heymans} et~al.}(2012)\citenamefont{{Heymans}, {Van
  Waerbeke}, {Miller}, {Erben}, {Hildebrandt}, {Hoekstra}, {Kitching},
  {Mellier}, {Simon}, {Bonnett} et~al.}}]{2012MNRAS.427..146H}
\bibinfo{author}{\bibfnamefont{C.}~\bibnamefont{{Heymans}}},
  \bibinfo{author}{\bibfnamefont{L.}~\bibnamefont{{Van Waerbeke}}},
  \bibinfo{author}{\bibfnamefont{L.}~\bibnamefont{{Miller}}},
  \bibinfo{author}{\bibfnamefont{T.}~\bibnamefont{{Erben}}},
  \bibinfo{author}{\bibfnamefont{H.}~\bibnamefont{{Hildebrandt}}},
  \bibinfo{author}{\bibfnamefont{H.}~\bibnamefont{{Hoekstra}}},
  \bibinfo{author}{\bibfnamefont{T.~D.} \bibnamefont{{Kitching}}},
  \bibinfo{author}{\bibfnamefont{Y.}~\bibnamefont{{Mellier}}},
  \bibinfo{author}{\bibfnamefont{P.}~\bibnamefont{{Simon}}},
  \bibinfo{author}{\bibfnamefont{C.}~\bibnamefont{{Bonnett}}},
  \bibnamefont{et~al.}, \bibinfo{journal}{\mnras}
  \textbf{\bibinfo{volume}{427}}, \bibinfo{pages}{146} (\bibinfo{year}{2012}),
  \eprint{1210.0032}.

\bibitem[{\citenamefont{{Kitching} et~al.}(2014)\citenamefont{{Kitching},
  {Heavens}, {Alsing}, {Erben}, {Heymans}, {Hildebrandt}, {Hoekstra}, {Jaffe},
  {Kiessling}, {Mellier} et~al.}}]{2014MNRAS.442.1326K}
\bibinfo{author}{\bibfnamefont{T.~D.} \bibnamefont{{Kitching}}},
  \bibinfo{author}{\bibfnamefont{A.~F.} \bibnamefont{{Heavens}}},
  \bibinfo{author}{\bibfnamefont{J.}~\bibnamefont{{Alsing}}},
  \bibinfo{author}{\bibfnamefont{T.}~\bibnamefont{{Erben}}},
  \bibinfo{author}{\bibfnamefont{C.}~\bibnamefont{{Heymans}}},
  \bibinfo{author}{\bibfnamefont{H.}~\bibnamefont{{Hildebrandt}}},
  \bibinfo{author}{\bibfnamefont{H.}~\bibnamefont{{Hoekstra}}},
  \bibinfo{author}{\bibfnamefont{A.}~\bibnamefont{{Jaffe}}},
  \bibinfo{author}{\bibfnamefont{A.}~\bibnamefont{{Kiessling}}},
  \bibinfo{author}{\bibfnamefont{Y.}~\bibnamefont{{Mellier}}},
  \bibnamefont{et~al.}, \bibinfo{journal}{\mnras}
  \textbf{\bibinfo{volume}{442}}, \bibinfo{pages}{1326} (\bibinfo{year}{2014}),
  \eprint{1401.6842}.

\bibitem[{\citenamefont{{Joudaki} et~al.}(2017)\citenamefont{{Joudaki},
  {Blake}, {Heymans}, {Choi}, {Harnois-Deraps}, {Hildebrandt}, {Joachimi},
  {Johnson}, {Mead}, {Parkinson} et~al.}}]{2017MNRAS.465.2033J}
\bibinfo{author}{\bibfnamefont{S.}~\bibnamefont{{Joudaki}}},
  \bibinfo{author}{\bibfnamefont{C.}~\bibnamefont{{Blake}}},
  \bibinfo{author}{\bibfnamefont{C.}~\bibnamefont{{Heymans}}},
  \bibinfo{author}{\bibfnamefont{A.}~\bibnamefont{{Choi}}},
  \bibinfo{author}{\bibfnamefont{J.}~\bibnamefont{{Harnois-Deraps}}},
  \bibinfo{author}{\bibfnamefont{H.}~\bibnamefont{{Hildebrandt}}},
  \bibinfo{author}{\bibfnamefont{B.}~\bibnamefont{{Joachimi}}},
  \bibinfo{author}{\bibfnamefont{A.}~\bibnamefont{{Johnson}}},
  \bibinfo{author}{\bibfnamefont{A.}~\bibnamefont{{Mead}}},
  \bibinfo{author}{\bibfnamefont{D.}~\bibnamefont{{Parkinson}}},
  \bibnamefont{et~al.}, \bibinfo{journal}{\mnras}
  \textbf{\bibinfo{volume}{465}}, \bibinfo{pages}{2033} (\bibinfo{year}{2017}),
  \eprint{1601.05786}.

\bibitem[{\citenamefont{{Zaldarriaga} and
  {Seljak}}(1997)}]{1997PhRvD..55.1830Z}
\bibinfo{author}{\bibfnamefont{M.}~\bibnamefont{{Zaldarriaga}}}
  \bibnamefont{and} \bibinfo{author}{\bibfnamefont{U.}~\bibnamefont{{Seljak}}},
  \bibinfo{journal}{\prd} \textbf{\bibinfo{volume}{55}}, \bibinfo{pages}{1830}
  (\bibinfo{year}{1997}), \eprint{astro-ph/9609170}.

\bibitem[{\citenamefont{{Kilbinger} et~al.}(2017)\citenamefont{{Kilbinger},
  {Heymans}, {Asgari}, {Joudaki}, {Schneider}, {Simon}, {Van Waerbeke},
  {Harnois-D{\'e}raps}, {Hildebrandt}, {K{\"o}hlinger}
  et~al.}}]{2017MNRAS.472.2126K}
\bibinfo{author}{\bibfnamefont{M.}~\bibnamefont{{Kilbinger}}},
  \bibinfo{author}{\bibfnamefont{C.}~\bibnamefont{{Heymans}}},
  \bibinfo{author}{\bibfnamefont{M.}~\bibnamefont{{Asgari}}},
  \bibinfo{author}{\bibfnamefont{S.}~\bibnamefont{{Joudaki}}},
  \bibinfo{author}{\bibfnamefont{P.}~\bibnamefont{{Schneider}}},
  \bibinfo{author}{\bibfnamefont{P.}~\bibnamefont{{Simon}}},
  \bibinfo{author}{\bibfnamefont{L.}~\bibnamefont{{Van Waerbeke}}},
  \bibinfo{author}{\bibfnamefont{J.}~\bibnamefont{{Harnois-D{\'e}raps}}},
  \bibinfo{author}{\bibfnamefont{H.}~\bibnamefont{{Hildebrandt}}},
  \bibinfo{author}{\bibfnamefont{F.}~\bibnamefont{{K{\"o}hlinger}}},
  \bibnamefont{et~al.}, \bibinfo{journal}{\mnras}
  \textbf{\bibinfo{volume}{472}}, \bibinfo{pages}{2126} (\bibinfo{year}{2017}),
  \eprint{1702.05301}.

\bibitem[{\citenamefont{{Limber}}(1953)}]{1953ApJ...117..134L}
\bibinfo{author}{\bibfnamefont{D.~N.} \bibnamefont{{Limber}}},
  \bibinfo{journal}{\apj} \textbf{\bibinfo{volume}{117}}, \bibinfo{pages}{134}
  (\bibinfo{year}{1953}).

\bibitem[{\citenamefont{{Afshordi} et~al.}(2004)\citenamefont{{Afshordi},
  {Loh}, and {Strauss}}}]{2004PhRvD..69h3524A}
\bibinfo{author}{\bibfnamefont{N.}~\bibnamefont{{Afshordi}}},
  \bibinfo{author}{\bibfnamefont{Y.-S.} \bibnamefont{{Loh}}}, \bibnamefont{and}
  \bibinfo{author}{\bibfnamefont{M.~A.} \bibnamefont{{Strauss}}},
  \bibinfo{journal}{\prd} \textbf{\bibinfo{volume}{69}}, \bibinfo{eid}{083524}
  (\bibinfo{year}{2004}), \eprint{astro-ph/0308260}.

\bibitem[{\citenamefont{{Chisari} et~al.}(2018)\citenamefont{{Chisari},
  {Alonso}, {Krause}, {Leonard}, {Bull}, {Neveu}, {Villarreal}, {Singh},
  {McClintock}, {Ellison} et~al.}}]{2018arXiv181205995C}
\bibinfo{author}{\bibfnamefont{N.~E.} \bibnamefont{{Chisari}}},
  \bibinfo{author}{\bibfnamefont{D.}~\bibnamefont{{Alonso}}},
  \bibinfo{author}{\bibfnamefont{E.}~\bibnamefont{{Krause}}},
  \bibinfo{author}{\bibfnamefont{C.~D.} \bibnamefont{{Leonard}}},
  \bibinfo{author}{\bibfnamefont{P.}~\bibnamefont{{Bull}}},
  \bibinfo{author}{\bibfnamefont{J.}~\bibnamefont{{Neveu}}},
  \bibinfo{author}{\bibfnamefont{A.}~\bibnamefont{{Villarreal}}},
  \bibinfo{author}{\bibfnamefont{S.}~\bibnamefont{{Singh}}},
  \bibinfo{author}{\bibfnamefont{T.}~\bibnamefont{{McClintock}}},
  \bibinfo{author}{\bibfnamefont{J.}~\bibnamefont{{Ellison}}},
  \bibnamefont{et~al.}, \bibinfo{journal}{arXiv e-prints}
  \bibinfo{eid}{arXiv:1812.05995} (\bibinfo{year}{2018}), \eprint{1812.05995}.

\bibitem[{\citenamefont{{Lesgourgues}}(2011)}]{2011arXiv1104.2932L}
\bibinfo{author}{\bibfnamefont{J.}~\bibnamefont{{Lesgourgues}}},
  \bibinfo{journal}{arXiv e-prints} \bibinfo{eid}{arXiv:1104.2932}
  (\bibinfo{year}{2011}), \eprint{1104.2932}.

\bibitem[{\citenamefont{{Takahashi} et~al.}(2012)\citenamefont{{Takahashi},
  {Sato}, {Nishimichi}, {Taruya}, and {Oguri}}}]{2012ApJ...761..152T}
\bibinfo{author}{\bibfnamefont{R.}~\bibnamefont{{Takahashi}}},
  \bibinfo{author}{\bibfnamefont{M.}~\bibnamefont{{Sato}}},
  \bibinfo{author}{\bibfnamefont{T.}~\bibnamefont{{Nishimichi}}},
  \bibinfo{author}{\bibfnamefont{A.}~\bibnamefont{{Taruya}}}, \bibnamefont{and}
  \bibinfo{author}{\bibfnamefont{M.}~\bibnamefont{{Oguri}}},
  \bibinfo{journal}{\apj} \textbf{\bibinfo{volume}{761}}, \bibinfo{eid}{152}
  (\bibinfo{year}{2012}), \eprint{1208.2701}.

\bibitem[{\citenamefont{{Tegmark} et~al.}(1997)\citenamefont{{Tegmark},
  {Taylor}, and {Heavens}}}]{1997ApJ...480...22T}
\bibinfo{author}{\bibfnamefont{M.}~\bibnamefont{{Tegmark}}},
  \bibinfo{author}{\bibfnamefont{A.~N.} \bibnamefont{{Taylor}}},
  \bibnamefont{and} \bibinfo{author}{\bibfnamefont{A.~F.}
  \bibnamefont{{Heavens}}}, \bibinfo{journal}{\apj}
  \textbf{\bibinfo{volume}{480}}, \bibinfo{pages}{22} (\bibinfo{year}{1997}),
  \eprint{astro-ph/9603021}.

\bibitem[{\citenamefont{{Bond}}(1995)}]{1995PhRvL..74.4369B}
\bibinfo{author}{\bibfnamefont{J.~R.} \bibnamefont{{Bond}}},
  \bibinfo{journal}{\prl} \textbf{\bibinfo{volume}{74}}, \bibinfo{pages}{4369}
  (\bibinfo{year}{1995}), \eprint{astro-ph/9407044}.

\bibitem[{\citenamefont{{Vogeley} and {Szalay}}(1996)}]{1996ApJ...465...34V}
\bibinfo{author}{\bibfnamefont{M.~S.} \bibnamefont{{Vogeley}}}
  \bibnamefont{and} \bibinfo{author}{\bibfnamefont{A.~S.}
  \bibnamefont{{Szalay}}}, \bibinfo{journal}{\apj}
  \textbf{\bibinfo{volume}{465}}, \bibinfo{pages}{34} (\bibinfo{year}{1996}),
  \eprint{astro-ph/9601185}.

\bibitem[{\citenamefont{{Erben} et~al.}(2013)\citenamefont{{Erben},
  {Hildebrandt}, {Miller}, {van Waerbeke}, {Heymans}, {Hoekstra}, {Kitching},
  {Mellier}, {Benjamin}, {Blake} et~al.}}]{2013MNRAS.433.2545E}
\bibinfo{author}{\bibfnamefont{T.}~\bibnamefont{{Erben}}},
  \bibinfo{author}{\bibfnamefont{H.}~\bibnamefont{{Hildebrandt}}},
  \bibinfo{author}{\bibfnamefont{L.}~\bibnamefont{{Miller}}},
  \bibinfo{author}{\bibfnamefont{L.}~\bibnamefont{{van Waerbeke}}},
  \bibinfo{author}{\bibfnamefont{C.}~\bibnamefont{{Heymans}}},
  \bibinfo{author}{\bibfnamefont{H.}~\bibnamefont{{Hoekstra}}},
  \bibinfo{author}{\bibfnamefont{T.~D.} \bibnamefont{{Kitching}}},
  \bibinfo{author}{\bibfnamefont{Y.}~\bibnamefont{{Mellier}}},
  \bibinfo{author}{\bibfnamefont{J.}~\bibnamefont{{Benjamin}}},
  \bibinfo{author}{\bibfnamefont{C.}~\bibnamefont{{Blake}}},
  \bibnamefont{et~al.}, \bibinfo{journal}{\mnras}
  \textbf{\bibinfo{volume}{433}}, \bibinfo{pages}{2545} (\bibinfo{year}{2013}),
  \eprint{1210.8156}.

\bibitem[{\citenamefont{{Miller} et~al.}(2013)\citenamefont{{Miller},
  {Heymans}, {Kitching}, {van Waerbeke}, {Erben}, {Hildebrandt}, {Hoekstra},
  {Mellier}, {Rowe}, {Coupon} et~al.}}]{2013MNRAS.429.2858M}
\bibinfo{author}{\bibfnamefont{L.}~\bibnamefont{{Miller}}},
  \bibinfo{author}{\bibfnamefont{C.}~\bibnamefont{{Heymans}}},
  \bibinfo{author}{\bibfnamefont{T.~D.} \bibnamefont{{Kitching}}},
  \bibinfo{author}{\bibfnamefont{L.}~\bibnamefont{{van Waerbeke}}},
  \bibinfo{author}{\bibfnamefont{T.}~\bibnamefont{{Erben}}},
  \bibinfo{author}{\bibfnamefont{H.}~\bibnamefont{{Hildebrandt}}},
  \bibinfo{author}{\bibfnamefont{H.}~\bibnamefont{{Hoekstra}}},
  \bibinfo{author}{\bibfnamefont{Y.}~\bibnamefont{{Mellier}}},
  \bibinfo{author}{\bibfnamefont{B.~T.~P.} \bibnamefont{{Rowe}}},
  \bibinfo{author}{\bibfnamefont{J.}~\bibnamefont{{Coupon}}},
  \bibnamefont{et~al.}, \bibinfo{journal}{\mnras}
  \textbf{\bibinfo{volume}{429}}, \bibinfo{pages}{2858} (\bibinfo{year}{2013}),
  \eprint{1210.8201}.

\bibitem[{\citenamefont{{Hildebrandt} et~al.}(2012)\citenamefont{{Hildebrandt},
  {Erben}, {Kuijken}, {van Waerbeke}, {Heymans}, {Coupon}, {Benjamin},
  {Bonnett}, {Fu}, {Hoekstra} et~al.}}]{2012MNRAS.421.2355H}
\bibinfo{author}{\bibfnamefont{H.}~\bibnamefont{{Hildebrandt}}},
  \bibinfo{author}{\bibfnamefont{T.}~\bibnamefont{{Erben}}},
  \bibinfo{author}{\bibfnamefont{K.}~\bibnamefont{{Kuijken}}},
  \bibinfo{author}{\bibfnamefont{L.}~\bibnamefont{{van Waerbeke}}},
  \bibinfo{author}{\bibfnamefont{C.}~\bibnamefont{{Heymans}}},
  \bibinfo{author}{\bibfnamefont{J.}~\bibnamefont{{Coupon}}},
  \bibinfo{author}{\bibfnamefont{J.}~\bibnamefont{{Benjamin}}},
  \bibinfo{author}{\bibfnamefont{C.}~\bibnamefont{{Bonnett}}},
  \bibinfo{author}{\bibfnamefont{L.}~\bibnamefont{{Fu}}},
  \bibinfo{author}{\bibfnamefont{H.}~\bibnamefont{{Hoekstra}}},
  \bibnamefont{et~al.}, \bibinfo{journal}{\mnras}
  \textbf{\bibinfo{volume}{421}}, \bibinfo{pages}{2355} (\bibinfo{year}{2012}),
  \eprint{1111.4434}.

\bibitem[{\citenamefont{{Benjamin} et~al.}(2013)\citenamefont{{Benjamin}, {Van
  Waerbeke}, {Heymans}, {Kilbinger}, {Erben}, {Hildebrandt}, {Hoekstra},
  {Kitching}, {Mellier}, {Miller} et~al.}}]{2013MNRAS.431.1547B}
\bibinfo{author}{\bibfnamefont{J.}~\bibnamefont{{Benjamin}}},
  \bibinfo{author}{\bibfnamefont{L.}~\bibnamefont{{Van Waerbeke}}},
  \bibinfo{author}{\bibfnamefont{C.}~\bibnamefont{{Heymans}}},
  \bibinfo{author}{\bibfnamefont{M.}~\bibnamefont{{Kilbinger}}},
  \bibinfo{author}{\bibfnamefont{T.}~\bibnamefont{{Erben}}},
  \bibinfo{author}{\bibfnamefont{H.}~\bibnamefont{{Hildebrandt}}},
  \bibinfo{author}{\bibfnamefont{H.}~\bibnamefont{{Hoekstra}}},
  \bibinfo{author}{\bibfnamefont{T.~D.} \bibnamefont{{Kitching}}},
  \bibinfo{author}{\bibfnamefont{Y.}~\bibnamefont{{Mellier}}},
  \bibinfo{author}{\bibfnamefont{L.}~\bibnamefont{{Miller}}},
  \bibnamefont{et~al.}, \bibinfo{journal}{\mnras}
  \textbf{\bibinfo{volume}{431}}, \bibinfo{pages}{1547} (\bibinfo{year}{2013}),
  \eprint{1212.3327}.

\bibitem[{\citenamefont{{Ben{\'{\i}}tez}}(2000)}]{2000ApJ...536..571B}
\bibinfo{author}{\bibfnamefont{N.}~\bibnamefont{{Ben{\'{\i}}tez}}},
  \bibinfo{journal}{\apj} \textbf{\bibinfo{volume}{536}}, \bibinfo{pages}{571}
  (\bibinfo{year}{2000}), \eprint{astro-ph/9811189}.

\bibitem[{\citenamefont{{Alonso} et~al.}(2018)\citenamefont{{Alonso},
  {Sanchez}, and {Slosar}}}]{2018arXiv180909603A}
\bibinfo{author}{\bibfnamefont{D.}~\bibnamefont{{Alonso}}},
  \bibinfo{author}{\bibfnamefont{J.}~\bibnamefont{{Sanchez}}},
  \bibnamefont{and} \bibinfo{author}{\bibfnamefont{A.}~\bibnamefont{{Slosar}}},
  \bibinfo{journal}{arXiv e-prints} \bibinfo{eid}{arXiv:1809.09603}
  (\bibinfo{year}{2018}), \eprint{1809.09603}.

\bibitem[{\citenamefont{{K{\"o}hlinger}
  et~al.}(2016)\citenamefont{{K{\"o}hlinger}, {Viola}, {Valkenburg},
  {Joachimi}, {Hoekstra}, and {Kuijken}}}]{2016MNRAS.456.1508K}
\bibinfo{author}{\bibfnamefont{F.}~\bibnamefont{{K{\"o}hlinger}}},
  \bibinfo{author}{\bibfnamefont{M.}~\bibnamefont{{Viola}}},
  \bibinfo{author}{\bibfnamefont{W.}~\bibnamefont{{Valkenburg}}},
  \bibinfo{author}{\bibfnamefont{B.}~\bibnamefont{{Joachimi}}},
  \bibinfo{author}{\bibfnamefont{H.}~\bibnamefont{{Hoekstra}}},
  \bibnamefont{and}
  \bibinfo{author}{\bibfnamefont{K.}~\bibnamefont{{Kuijken}}},
  \bibinfo{journal}{\mnras} \textbf{\bibinfo{volume}{456}},
  \bibinfo{pages}{1508} (\bibinfo{year}{2016}), \eprint{1509.04071}.

\bibitem[{\citenamefont{{Asgari} et~al.}(2017)\citenamefont{{Asgari},
  {Heymans}, {Blake}, {Harnois-Deraps}, {Schneider}, and {Van
  Waerbeke}}}]{2017MNRAS.464.1676A}
\bibinfo{author}{\bibfnamefont{M.}~\bibnamefont{{Asgari}}},
  \bibinfo{author}{\bibfnamefont{C.}~\bibnamefont{{Heymans}}},
  \bibinfo{author}{\bibfnamefont{C.}~\bibnamefont{{Blake}}},
  \bibinfo{author}{\bibfnamefont{J.}~\bibnamefont{{Harnois-Deraps}}},
  \bibinfo{author}{\bibfnamefont{P.}~\bibnamefont{{Schneider}}},
  \bibnamefont{and} \bibinfo{author}{\bibfnamefont{L.}~\bibnamefont{{Van
  Waerbeke}}}, \bibinfo{journal}{\mnras} \textbf{\bibinfo{volume}{464}},
  \bibinfo{pages}{1676} (\bibinfo{year}{2017}), \eprint{1601.00115}.

\bibitem[{\citenamefont{{Alsing} et~al.}(2017)\citenamefont{{Alsing},
  {Heavens}, and {Jaffe}}}]{2017MNRAS.466.3272A}
\bibinfo{author}{\bibfnamefont{J.}~\bibnamefont{{Alsing}}},
  \bibinfo{author}{\bibfnamefont{A.}~\bibnamefont{{Heavens}}},
  \bibnamefont{and} \bibinfo{author}{\bibfnamefont{A.~H.}
  \bibnamefont{{Jaffe}}}, \bibinfo{journal}{\mnras}
  \textbf{\bibinfo{volume}{466}}, \bibinfo{pages}{3272} (\bibinfo{year}{2017}),
  \eprint{1607.00008}.

\bibitem[{\citenamefont{{Kilbinger} et~al.}(2014)\citenamefont{{Kilbinger},
  {Bonnett}, and {Coupon}}}]{2014ascl.soft02026K}
\bibinfo{author}{\bibfnamefont{M.}~\bibnamefont{{Kilbinger}}},
  \bibinfo{author}{\bibfnamefont{C.}~\bibnamefont{{Bonnett}}},
  \bibnamefont{and} \bibinfo{author}{\bibfnamefont{J.}~\bibnamefont{{Coupon}}},
  \emph{\bibinfo{title}{{athena: Tree code for second-order correlation
  functions}}}, \bibinfo{howpublished}{Astrophysics Source Code Library}
  (\bibinfo{year}{2014}), \eprint{1402.026}.

\bibitem[{\citenamefont{{Hartlap} et~al.}(2007)\citenamefont{{Hartlap},
  {Simon}, and {Schneider}}}]{2007A&A...464..399H}
\bibinfo{author}{\bibfnamefont{J.}~\bibnamefont{{Hartlap}}},
  \bibinfo{author}{\bibfnamefont{P.}~\bibnamefont{{Simon}}}, \bibnamefont{and}
  \bibinfo{author}{\bibfnamefont{P.}~\bibnamefont{{Schneider}}},
  \bibinfo{journal}{\aap} \textbf{\bibinfo{volume}{464}}, \bibinfo{pages}{399}
  (\bibinfo{year}{2007}), \eprint{astro-ph/0608064}.

\bibitem[{\citenamefont{{Sellentin} and {Heavens}}(2016)}]{2016MNRAS.456L.132S}
\bibinfo{author}{\bibfnamefont{E.}~\bibnamefont{{Sellentin}}} \bibnamefont{and}
  \bibinfo{author}{\bibfnamefont{A.~F.} \bibnamefont{{Heavens}}},
  \bibinfo{journal}{\mnras} \textbf{\bibinfo{volume}{456}},
  \bibinfo{pages}{L132} (\bibinfo{year}{2016}), \eprint{1511.05969}.

\bibitem[{\citenamefont{{Barreira} et~al.}(2018)\citenamefont{{Barreira},
  {Krause}, and {Schmidt}}}]{2018JCAP...10..053B}
\bibinfo{author}{\bibfnamefont{A.}~\bibnamefont{{Barreira}}},
  \bibinfo{author}{\bibfnamefont{E.}~\bibnamefont{{Krause}}}, \bibnamefont{and}
  \bibinfo{author}{\bibfnamefont{F.}~\bibnamefont{{Schmidt}}},
  \bibinfo{journal}{Journal of Cosmology and Astro-Particle Physics}
  \textbf{\bibinfo{volume}{2018}}, \bibinfo{eid}{053} (\bibinfo{year}{2018}),
  \eprint{1807.04266}.

\bibitem[{\citenamefont{{Harnois-D{\'e}raps} and {van
  Waerbeke}}(2015)}]{2015MNRAS.450.2857H}
\bibinfo{author}{\bibfnamefont{J.}~\bibnamefont{{Harnois-D{\'e}raps}}}
  \bibnamefont{and} \bibinfo{author}{\bibfnamefont{L.}~\bibnamefont{{van
  Waerbeke}}}, \bibinfo{journal}{\mnras} \textbf{\bibinfo{volume}{450}},
  \bibinfo{pages}{2857} (\bibinfo{year}{2015}), \eprint{1406.0543}.

\bibitem[{\citenamefont{{Foreman-Mackey}
  et~al.}(2013)\citenamefont{{Foreman-Mackey}, {Hogg}, {Lang}, and
  {Goodman}}}]{2013PASP..125..306F}
\bibinfo{author}{\bibfnamefont{D.}~\bibnamefont{{Foreman-Mackey}}},
  \bibinfo{author}{\bibfnamefont{D.~W.} \bibnamefont{{Hogg}}},
  \bibinfo{author}{\bibfnamefont{D.}~\bibnamefont{{Lang}}}, \bibnamefont{and}
  \bibinfo{author}{\bibfnamefont{J.}~\bibnamefont{{Goodman}}},
  \bibinfo{journal}{\pasp} \textbf{\bibinfo{volume}{125}}, \bibinfo{pages}{306}
  (\bibinfo{year}{2013}), \eprint{1202.3665}.

\bibitem[{\citenamefont{Audren et~al.}(2013)\citenamefont{Audren, Lesgourgues,
  Benabed, and Prunet}}]{Audren:2012wb}
\bibinfo{author}{\bibfnamefont{B.}~\bibnamefont{Audren}},
  \bibinfo{author}{\bibfnamefont{J.}~\bibnamefont{Lesgourgues}},
  \bibinfo{author}{\bibfnamefont{K.}~\bibnamefont{Benabed}}, \bibnamefont{and}
  \bibinfo{author}{\bibfnamefont{S.}~\bibnamefont{Prunet}},
  \bibinfo{journal}{JCAP} \textbf{\bibinfo{volume}{1302}}, \bibinfo{pages}{001}
  (\bibinfo{year}{2013}), \eprint{1210.7183}.

\bibitem[{\citenamefont{Brinckmann and Lesgourgues}(2018)}]{Brinckmann:2018cvx}
\bibinfo{author}{\bibfnamefont{T.}~\bibnamefont{Brinckmann}} \bibnamefont{and}
  \bibinfo{author}{\bibfnamefont{J.}~\bibnamefont{Lesgourgues}}
  (\bibinfo{year}{2018}), \eprint{1804.07261}.

\bibitem[{\citenamefont{{Sellentin} and {Heavens}}(2018)}]{2018MNRAS.473.2355S}
\bibinfo{author}{\bibfnamefont{E.}~\bibnamefont{{Sellentin}}} \bibnamefont{and}
  \bibinfo{author}{\bibfnamefont{A.~F.} \bibnamefont{{Heavens}}},
  \bibinfo{journal}{\mnras} \textbf{\bibinfo{volume}{473}},
  \bibinfo{pages}{2355} (\bibinfo{year}{2018}), \eprint{1707.04488}.

\bibitem[{\citenamefont{{Hamimeche} and {Lewis}}(2008)}]{2008PhRvD..77j3013H}
\bibinfo{author}{\bibfnamefont{S.}~\bibnamefont{{Hamimeche}}} \bibnamefont{and}
  \bibinfo{author}{\bibfnamefont{A.}~\bibnamefont{{Lewis}}},
  \bibinfo{journal}{\prd} \textbf{\bibinfo{volume}{77}}, \bibinfo{eid}{103013}
  (\bibinfo{year}{2008}), \eprint{0801.0554}.

\bibitem[{\citenamefont{{Chevallier} and
  {Polarski}}(2001)}]{2001IJMPD..10..213C}
\bibinfo{author}{\bibfnamefont{M.}~\bibnamefont{{Chevallier}}}
  \bibnamefont{and}
  \bibinfo{author}{\bibfnamefont{D.}~\bibnamefont{{Polarski}}},
  \bibinfo{journal}{International Journal of Modern Physics D}
  \textbf{\bibinfo{volume}{10}}, \bibinfo{pages}{213} (\bibinfo{year}{2001}),
  \eprint{gr-qc/0009008}.

\bibitem[{\citenamefont{{Simon} et~al.}(2015)\citenamefont{{Simon},
  {Semboloni}, {van Waerbeke}, {Hoekstra}, {Erben}, {Fu}, {Harnois-D{\'e}raps},
  {Heymans}, {Hildebrandt}, {Kilbinger} et~al.}}]{2015MNRAS.449.1505S}
\bibinfo{author}{\bibfnamefont{P.}~\bibnamefont{{Simon}}},
  \bibinfo{author}{\bibfnamefont{E.}~\bibnamefont{{Semboloni}}},
  \bibinfo{author}{\bibfnamefont{L.}~\bibnamefont{{van Waerbeke}}},
  \bibinfo{author}{\bibfnamefont{H.}~\bibnamefont{{Hoekstra}}},
  \bibinfo{author}{\bibfnamefont{T.}~\bibnamefont{{Erben}}},
  \bibinfo{author}{\bibfnamefont{L.}~\bibnamefont{{Fu}}},
  \bibinfo{author}{\bibfnamefont{J.}~\bibnamefont{{Harnois-D{\'e}raps}}},
  \bibinfo{author}{\bibfnamefont{C.}~\bibnamefont{{Heymans}}},
  \bibinfo{author}{\bibfnamefont{H.}~\bibnamefont{{Hildebrandt}}},
  \bibinfo{author}{\bibfnamefont{M.}~\bibnamefont{{Kilbinger}}},
  \bibnamefont{et~al.}, \bibinfo{journal}{\mnras}
  \textbf{\bibinfo{volume}{449}}, \bibinfo{pages}{1505} (\bibinfo{year}{2015}),
  \eprint{1502.04575}.

\bibitem[{\citenamefont{{Troxel} and {Ishak}}(2015)}]{2015PhR...558....1T}
\bibinfo{author}{\bibfnamefont{M.~A.} \bibnamefont{{Troxel}}} \bibnamefont{and}
  \bibinfo{author}{\bibfnamefont{M.}~\bibnamefont{{Ishak}}},
  \bibinfo{journal}{\physrep} \textbf{\bibinfo{volume}{558}},
  \bibinfo{pages}{1} (\bibinfo{year}{2015}), \eprint{1407.6990}.

\bibitem[{\citenamefont{van Uitert et~al.}(2018)}]{vanUitert:2017ieu}
\bibinfo{author}{\bibfnamefont{E.}~\bibnamefont{van Uitert}}
  \bibnamefont{et~al.}, \bibinfo{journal}{Mon. Not. Roy. Astron. Soc.}
  \textbf{\bibinfo{volume}{476}}, \bibinfo{pages}{4662} (\bibinfo{year}{2018}),
  \eprint{1706.05004}.

\bibitem[{\citenamefont{Joudaki et~al.}(2018)}]{Joudaki:2017zdt}
\bibinfo{author}{\bibfnamefont{S.}~\bibnamefont{Joudaki}} \bibnamefont{et~al.},
  \bibinfo{journal}{Mon. Not. Roy. Astron. Soc.}
  \textbf{\bibinfo{volume}{474}}, \bibinfo{pages}{4894} (\bibinfo{year}{2018}),
  \eprint{1707.06627}.

\bibitem[{\citenamefont{{Abbott} et~al.}(2018)\citenamefont{{Abbott},
  {Abdalla}, {Alarcon}, {Aleksi{\'c}}, {Allam}, {Allen}, {Amara}, {Annis},
  {Asorey}, {Avila} et~al.}}]{2018PhRvD..98d3526A}
\bibinfo{author}{\bibfnamefont{T.~M.~C.} \bibnamefont{{Abbott}}},
  \bibinfo{author}{\bibfnamefont{F.~B.} \bibnamefont{{Abdalla}}},
  \bibinfo{author}{\bibfnamefont{A.}~\bibnamefont{{Alarcon}}},
  \bibinfo{author}{\bibfnamefont{J.}~\bibnamefont{{Aleksi{\'c}}}},
  \bibinfo{author}{\bibfnamefont{S.}~\bibnamefont{{Allam}}},
  \bibinfo{author}{\bibfnamefont{S.}~\bibnamefont{{Allen}}},
  \bibinfo{author}{\bibfnamefont{A.}~\bibnamefont{{Amara}}},
  \bibinfo{author}{\bibfnamefont{J.}~\bibnamefont{{Annis}}},
  \bibinfo{author}{\bibfnamefont{J.}~\bibnamefont{{Asorey}}},
  \bibinfo{author}{\bibfnamefont{S.}~\bibnamefont{{Avila}}},
  \bibnamefont{et~al.}, \bibinfo{journal}{\prd} \textbf{\bibinfo{volume}{98}},
  \bibinfo{eid}{043526} (\bibinfo{year}{2018}), \eprint{1708.01530}.

\bibitem[{\citenamefont{{Reichardt} et~al.}(2001)\citenamefont{{Reichardt},
  {Jimenez}, and {Heavens}}}]{2001MNRAS.327..849R}
\bibinfo{author}{\bibfnamefont{C.}~\bibnamefont{{Reichardt}}},
  \bibinfo{author}{\bibfnamefont{R.}~\bibnamefont{{Jimenez}}},
  \bibnamefont{and} \bibinfo{author}{\bibfnamefont{A.~F.}
  \bibnamefont{{Heavens}}}, \bibinfo{journal}{\mnras}
  \textbf{\bibinfo{volume}{327}}, \bibinfo{pages}{849} (\bibinfo{year}{2001}),
  \eprint{astro-ph/0101074}.

\bibitem[{\citenamefont{{Alsing} and {Wandelt}}(2018)}]{2018MNRAS.476L..60A}
\bibinfo{author}{\bibfnamefont{J.}~\bibnamefont{{Alsing}}} \bibnamefont{and}
  \bibinfo{author}{\bibfnamefont{B.}~\bibnamefont{{Wandelt}}},
  \bibinfo{journal}{\mnras} \textbf{\bibinfo{volume}{476}},
  \bibinfo{pages}{L60} (\bibinfo{year}{2018}), \eprint{1712.00012}.

\bibitem[{\citenamefont{{Alsing} et~al.}(2018)\citenamefont{{Alsing},
  {Wandelt}, and {Feeney}}}]{2018MNRAS.477.2874A}
\bibinfo{author}{\bibfnamefont{J.}~\bibnamefont{{Alsing}}},
  \bibinfo{author}{\bibfnamefont{B.}~\bibnamefont{{Wandelt}}},
  \bibnamefont{and} \bibinfo{author}{\bibfnamefont{S.}~\bibnamefont{{Feeney}}},
  \bibinfo{journal}{\mnras} \textbf{\bibinfo{volume}{477}},
  \bibinfo{pages}{2874} (\bibinfo{year}{2018}), \eprint{1801.01497}.

\end{thebibliography}

\end{document}